\documentclass[11pt]{article}
\renewcommand{\theequation}{\arabic{section}.\arabic{equation}}

\newtheorem{example}{Example}[section]
\newtheorem{theorem}{Theorem}[section]
\newtheorem{corollary}{Corollary}[section]
\usepackage{bm}
\usepackage{longtable}
\usepackage{amssymb}
\usepackage{amsmath}
\usepackage{epsfig}

\begin{document}
\begin{center}
{\LARGE\bf Expansions about the gamma for the}\\[1ex]
{\LARGE\bf distribution and quantiles of a standard estimate}\\[1ex]
by\\[1ex]
Christopher S. Withers\\
Applied Mathematics Group\\
Industrial Research Limited\\
Lower Hutt, NEW ZEALAND\\[2ex]
Saralees Nadarajah\\
School of Mathematics\\
University of Manchester\\
Manchester M13 9PL, UK
\end{center}
\vspace{1.5cm}
{\bf Abstract:}~~We give expansions for the distribution, density,
and quantiles of an estimate, building on results of
Cornish, Fisher, Hill, Davis and the authors.
The estimate is assumed to be non-lattice with the  standard expansions for its cumulants.
By expanding about a skew variable
with matched skewness, one can drastically reduce the number of terms needed
for a given level of accuracy.
The building blocks generalize the Hermite polynomials.
We demonstrate with expansions about the gamma.


\section{Introduction and summary}
\setcounter{equation}{0}

The Cornish Fisher expansions due to
Cornish and Fisher (1937) and  Hill and Davis (1968)
have received applications in many areas of statistics.
The Cornish Fisher expansions also have applications in many applied areas, including
risk measures for hedge funds,
margin setting of index futures,
structural equation models,
modified sudden death tests,
blind inversion of Wiener systems,
GPS positioning accuracy estimation,
steady-state simulation analysis,
blind separation of post-nonlinear mixtures,
cycle time quantile estimation,
estimation of the maximum average time to flower,
performance of Skart,
testing and evaluation, load flow in systems with wind generation,
Value-at-Risk portfolio optimization,
quantile mechanics,
channel capacity in communications theory, economics, financial intermediation and physics.
Three of the most recent papers applying Cornish Fisher expansions to these areas are:
Alfredo and Arunachalam (2011), Simonato (2011) and Zhang {\it et al.} (2011).

The aim of this paper is to develop technical tools so that
the Cornish Fisher expansions could have wider applications.
In particular, we show how one can drastically reduce the number of terms needed
for a given level of accuracy.
The building blocks involve Bell polynomials and Hermite polynomials.
In-built routines for these polynomials are available in most computer algebra packages.

Let $\theta$ be an unknown real parameter with a non-lattice estimate
$\widehat{\theta}$ having the standard asymptotic cumulant expansions
\begin{eqnarray}
\kappa_r \left(\widehat{\theta}\right) \approx \sum_{i=r-1}^\infty a_{ri\theta}\ n^{-i},
\
r\geq 1,
\label{stand}
\end{eqnarray}
where $a_{10\theta}=\theta$, each $a_{ri\theta}$ is bounded in $n$, and $a_{21\theta}$ is bounded away from zero as $n$ increases.
We call such an estimate {\it a standard estimate}.
For example, a smooth function of a sample
mean is a standard estimate - see Withers (1983).
Formulas for the leading coefficients were given for parametric estimates in Withers (1982)
and for non-parametric estimates in Withers (1983, 1988).
The standardized form
\begin{eqnarray}
Y_{01\theta}= \left(n/a_{21\theta}\right)^{1/2}\ \left(\widehat{\theta}-\theta\right)
\label{stan}
\end{eqnarray}
has cumulants expandable as
\begin{eqnarray}
\kappa_r \left(Y_{01\theta}\right) \approx n^{r/2}\sum_{i=r-1}^\infty A_{ri\theta}\ n^{-i},
\
r\geq 1,
\label{riY}
\end{eqnarray}
where
\begin{eqnarray}
A_{10\theta} =0,
\
A_{ri\theta}= a_{ri\theta}/a_{21\theta}^{r/2}
\mbox{ for }(r,i)\neq (1,0).
\label{Arith}
\end{eqnarray}
Set $P_n(x) = \Pr (Y\leq x)$ for $Y=Y_{01\theta}$,
and denote its density by $p_n(x)$.
Cornish and Fisher (1937), and Fisher and Cornish (1960) gave expansions for $P_n(x)$ and its inverse.
These can be re-written in the form
\begin{eqnarray}
&&
P_n(x)    \approx  P(x) - p(x)\sum_{r=1}^\infty n^{-r/2}\ h_r(x,L),
\label{h}
\\
&&
P^{-1} \left(P_n(x)\right)   \approx  x -\sum_{r=1}^\infty n^{-r/2}\ f_r(x,L),
\label{f}
\\
&&
P^{-1}_n \left(P(x)\right)   \approx  x +\sum_{r=1}^\infty n^{-r/2}\ g_r(x,L),
\label{g}
\end{eqnarray}
where  $P=\Phi$ and $p=\phi$ are the distribution and density of a
standard normal variable, denoted $N\sim {\cal N}(0,1)$,
$e_r(x,L)$ is a polynomial in both $x$ and $L=(L_1,L_2, \ldots)$  for $e=h,f,g$,
\begin{eqnarray}
L_{r} =l_r/r!,
\
l_{r}  \approx \sum_{j=0}^\infty A_{r,r+j-\delta,\theta}\ n^{-j},
\
\delta  =I(r\geq 3),
\label{r3}
\end{eqnarray}
and $I(\cdot)$ is the indicator function.
However, they gave no indication of how to truncate these expansions for each adjusted cumulant $l_r$.
This was remedied in Withers (1984) which showed:

\begin{theorem}
With notation as above,
\begin{eqnarray}
&&
P_n(x)   \approx P(x) -p(x)\sum_{r=1}^\infty n^{-r/2}\ h_r(x),
\label{hhh}
\\
&&
P^{-1} \left(P_n(x)\right) \approx x -\sum_{r=1}^\infty n^{-r/2}\ f_r(x),
\label{fff}
\\
&&
P^{-1}_n \left(P(x)\right) \approx x +\sum_{r=1}^\infty n^{-r/2}\ g_r(x)
\label{ggg}
\end{eqnarray}
for $e=h$, $f$, $g$, where $e_r(x)$ is a polynomial in both $x$ and $A=\{A_{ri}\}$, $A_{ri}=A_{ri\theta}$,
and again, $P=\Phi$, $p=\phi$.
The cumulant coefficients needed for $e_{r}(x)$, $(e=h, f, g)$ are as follows:
\begin{eqnarray*}
\mbox{for} & r=1: & A_{11}\ \ \ \ \ \ A_{32};
\\
\mbox{for} & r=2: & \ \ \ \ A_{22}\ \ \ \ \ \ A_{43};
\\
\mbox{for} & r=3: & A_{12}\ \ \ \ \ \  A_{33}\ \ \ \ \ \ A_{54};
\\
\mbox{for} & r=4: & \ \ \ \  A_{23}\ \ \ \ \ \ A_{44}\ \ \ \ \ \ A_{65};
\\
\mbox{for} & r=5: & A_{13}\ \ \ \ \ \ A_{34}\ \ \ \ \ \ A_{55}\ \ \ \ \ \ A_{76};
\\
\mbox{for} & r=6: & \ \ \ \ A_{24}\ \ \ \ \ \ A_{45}\ \ \ \ \ \ A_{66}\ \ \ \ \ \ A_{87}.
\end{eqnarray*}
\end{theorem}

\noindent
{\bf Proof:}
The expansions  (\ref{hhh})-(\ref{ggg}) are obtained
by substituting $l_r$ of (\ref{r3}) into $e_r(x,L)$ of (\ref{h})-(\ref{g}), giving
\begin{eqnarray}
&&
e_r(x,L) \approx \sum_{i=0}^\infty e_{ri}(x) n^{-i} \mbox{ say},
\label{erxL}
\\
&&
e_{r}(x) = \sum_{0\leq i<r/2} e_{r-2i,i}(x) = e_{r} \left(x, L_0 \right) + \Delta_{re} \mbox{ say},
\label{3.1}
\end{eqnarray}
where
\begin{eqnarray}
&&
e_{r0}(x) = e_{r} \left(x, L_0 \right),
\
\Delta_{re}= \sum_{1\leq i<r/2} e_{r-2i,i}(x),
\label{3.2}
\\
&&
L_0 = \left(L_{01}, L_{02}, \ldots \right),
\
L_{0r}=l_{0r}/r!,
\
l_{0r}=A_{r,r-\delta,\theta},
\label{ll}
\end{eqnarray}
the leading term of $l_r$.
\
$\Box$

To apply the expansions of Theorem 1.1, one can calculate $\Phi(x)$ and its inverse using
say NAG routines G01EAF and G01FAF: see {\sf http:// www. nag. co. uk/ numeric/ numerical\_libraries.asp}
(Column 1 of Table II of  Cornish and Fisher (1937, page 318) gave the main
quantiles of $N$ to nine decimal places.)
By (\ref{ggg}), $\widehat{\theta}$ has $P(x)$-quantile
\begin{eqnarray}
\theta + \left( a_{21\theta}/n \right)^{1/2}\sum_{r=0}^\infty n^{-r/2}\ g_r(x),
\label{quan}
\end{eqnarray}
where $g_0(x)=x$.
A drawback of the method is the increasingly large number of terms in each $e_r(x)$ as $r$ increases.
The chief contributor is the skewness coefficient $l_{03}=A_{32}$,
followed by the bias coefficient  $l_{01}=A_{11}$ and  the second order variance
coefficient $l_{02}=A_{22}$.
We now show how to remove these terms.
Except for sample means, $\mathbb{E}\ \widehat{\theta}$ and
$var( \widehat{\theta})$ are unknown, but they can be approximated by truncating their cumulant expansions.
Set
\begin{eqnarray}
&&
s_{rJ\theta} = \sum_{i=r-1}^J a_{ri\theta} n^{-i},
\
J\geq r-1,
\label{rJ}
\\
&&
Y_{JK\theta} = s_{2K\theta}^{-1/2} \left(\widehat{\theta}-s_{1J\theta}\right),
\
J\geq 0,
K\geq 1.
\label{JKth}
\end{eqnarray}
Then,
\begin{eqnarray}
\kappa_r \left(Y_{JK\theta}\right) \approx n^{r/2}\sum_{i=r-1}^\infty A_{ri\theta}^{JK} n^{-i},
\
r\geq 1,
\label{rith}
\end{eqnarray}
where $A_{1i\theta}^{JK} = 0$ for $i\leq J$, $A_{2i\theta}^{JK}=0$ for $i\leq K$,
and the other  $A_{ri\theta}^{JK}$ are given in terms of  $\{A_{ri\theta}\}$ of (\ref{Arith}) in Section 6.
(For $r\geq 2$, $A_{ri\theta}^{JK}$ does not depend on $J$.)
Suppose that $a_{32\theta}>0$.
(If  $a_{32\theta}<0$, replace $(\widehat{\theta}, \theta)$ by  $(-\widehat{\theta}, -\theta)$.)

Now let $\widehat{w}$ be another non-lattice standard estimate with the standard cumulant expansion:
\begin{eqnarray*}
\kappa_r \left(\widehat{w}\right) \approx \sum_{i=r-1}^\infty a_{riw}\ m^{-i},
\
r\geq 1,
\end{eqnarray*}
where $a_{10w}=w$.
Since $m$ is arbitrary up to a multiplier, we can take $m=n\tau$ for some constant $\tau>0$.
We also assume that $a_{32w}>0$.
(If $a_{32w}<0$, replace $(\widehat{w},w)$ by  $(-\widehat{w},-w)$.)
By Theorem 6.1 below,
\begin{eqnarray}
\kappa_r \left(Y_{JK\theta}\right) - \kappa_r \left(Y_{JKw}\right)
\approx n^{r/2}\sum_{i=r-1}^\infty  A_{ri}^{JK} n^{-i},
\nonumber
\end{eqnarray}
where
\begin{eqnarray}
A_{ri}^{JK} = A_{ri\theta}^{JK}-\widetilde{A}_{riw}^{JK},
\
\widetilde{A}_{riw}^{JK}=\tau^{r/2-i} A_{riw}^{JK}.
\label{1.22}
\end{eqnarray}
By Withers and Nadarajah (2012), the expansions (\ref{h})-(\ref{g}), (\ref{hhh})-(\ref{ggg}) remain true for
\begin{eqnarray}
P_n(x)=\Pr(Y\leq x),
\
P(x)=\Pr(X\leq x),
\label{now1}
\end{eqnarray}
and $p_X(x)=p(x)$ is the density of $X$, if
\begin{eqnarray}
\kappa_r(Y)-\kappa_r(X)\approx n^{r/2}\sum_{i=r-1}^\infty  A_{ri} n^{-i},
\label{general}
\end{eqnarray}
and $X$, $Y$ are non-lattice.
So, these expansions hold for
\begin{eqnarray}
Y=Y_{JK\theta},
\
X=Y_{JKw},
\
{A}_{ri} =  A_{ri}^{JK}.
\label{now}
\end{eqnarray}
We assume that $w$ and the cumulant coefficients $a_{riw}$ are known.
To apply these expansions, we need to be able to calculate $P(x)$ and its
inverse accurately.
For $X$ linear in gamma, this can be done using NAG routines G01EFF and G01FFF.

Now choose $\tau$ so that $A_{32}^{JK}=0$, that is,
\begin{eqnarray}
\tau = \left({A}_{32w}/{A}_{32\theta}\right)^2.
\label{c3}
\end{eqnarray}
This has the effect of roughly halving the number of terms in each $e_r(x)$.
Table 1.1 compares the
number of terms in $e_r(x)$ for different choices of $J$, $K$.
The number is written as $N+M$, where  $N$ is the number of terms in
$e_r(x,L_0)$ and $M$ is the number of terms in $\Delta_{re}(x)$ of (\ref{3.2}).
For example, the line $e_0$ refers to the  approximation
\begin{eqnarray*}
\left(n/a_{21\theta}\right)^{1/2} \left(\widehat{\theta}-\theta\right) \stackrel{\cal L}{=}
\left(m/a_{21w}\right)^{1/2} \left(\widehat{w}-w\right),
\
m=n\tau,
\end{eqnarray*}
with $\tau$ of (\ref{c3}).
The four columns after the columns headed $J$ and $K$, accumulate the number of terms needed.
That is, they give the number of terms needed for error $O(n^{-(r+1)/2})$
when  (\ref{hhh})-(\ref{ggg}) are truncated after $r+1$ terms.
The final column gives the percentage savings in number of terms over
the Cornish-Fisher expansion, which amounts to choosing $\widehat{w}\sim {\cal N}(0,n^{-1})$.

For example,  to calculate a quantile of $\widehat{\theta}$ to $O(n^{-7/2})$ via $P_n^{-1}$
of (\ref{ggg}),  the Cornish-Fisher method
requires calculating 48+29=77 terms, but using $(J=3,K=4)$ requires only 11+7=18
terms - a saving of 77 percent.
Similarly,  to calculate a $p$-value
of $\widehat{\theta}$ to $O(n^{-2})$ via $P_n$ of (\ref{fff}),  the Cornish-Fisher
method
requires calculating 14+2=16 terms, but using $J=K=2$ requires only 3+1=4
terms  - a saving of 75 percent.
The percentage savings increases with $r$.

\noindent
[Table 1.1 about here.]

We draw attention to the choice $J=K=1$, since then $e_1(x)=0$,
and the gamma approximation has error only  $O(n^{-1})$, not
the usual $O(n^{-1/2})$.
This is the simple approximation
\begin{eqnarray*}
Y_{11\theta}=\left(n/a_{21\theta}\right)^{1/2} \left(\widehat{\theta}-\theta-a_{11\theta}n^{-1}\right)
\stackrel{\cal L}{=}
\left(m/a_{21w}\right)^{1/2} \left(\widehat{w}-w-a_{11w}m^{-1}\right),
\
m=n\tau
\end{eqnarray*}
with $\tau$ of (\ref{c3}).

Increasing $J$ or $K$ beyond those given in the table,
does not decrease the number of terms in $e_r(x)$.
However, if we want to calculate  $e_r(x)$ up to say $r=6$, then we should
choose $J$, $K$ as given on the line for  $r=6$, that is, $J=3$, $K=4$.

We now give  $e_r(x)$ up to $r=6$  in terms of  {\it the generalized Hermite polynomial},
\begin{eqnarray}
H_r=H_{rX}(x) =  p(x)^{-1} (-D)^r p(x),
\
D=d/dx,
\
r\geq 0.
\label{notHe}
\end{eqnarray}
$H_r=H_{rX}(x)$ may not be a polynomial in $x$, but it {\it is} a polynomial
in ${\bf a}=(a_1,a_2,\ldots)$, where
\begin{eqnarray}
a_r=a_{rX}=D^r a(x),
\
a(x)= a_X(x)=-\ln p(x),
\
r\geq 1,
\label{ar}
\end{eqnarray}
with generating function
\begin{eqnarray}
A_{X}(x,t)=\sum_{r=0}^\infty a_{rX} t^r/r!=a_X(x+t)=-\ln p_X(x+t).
\label{agen}
\end{eqnarray}
So, $a_1=H_1$ and $a(x)$ is convex if and only if $a_2\geq 0$.
The function $a_r$ is simpler to compute than $H_r$, (only the first two are
non-zero for the normal), but expressions in terms of ${\bf a}$
are longer (often much longer) than expressions in terms of ${\bf H}$.
Also $H_r$ is easily found from its generating function
\begin{eqnarray}
B_{X}(x,t) =\sum_{r=0}^\infty H_{rX}(x) t^r/r!=p_X(x-t)/p_X(x).
\label{Hgen}
\end{eqnarray}
For $e=h,f,g$, Theorem 1.2 gives the much simpler form for $e_r(x)$ of (\ref{3.1}).

\begin{theorem}
With notation as above,
\begin{eqnarray}
e_r(x) =\nabla_r+\nabla_{re}
\nonumber
\end{eqnarray}
for $e = h, f, g$, where
\begin{eqnarray}
&&
\nabla_r=\sum_{(r+1)/2\leq i\leq r+1} \overline{A}_{2i-r,i} H_{2i-r-1},
\nonumber
\\
&&
\overline{A}_{ri} = A_{ri}/r!.
\label{bar}
\end{eqnarray}
In particular,
\begin{eqnarray}
&&
\nabla_1 = 0\mbox{ if }J\geq 1,
\nonumber
\\
&&
\nabla_2 = \overline{A}_{43}H_2\mbox{ if }K\geq 2,
\nonumber
\\
&&
\nabla_3 = \overline{A}_{33}H_2 +\overline{A}_{54}H_4\mbox{ if }J\geq 2,
\nonumber
\\
&&
\nabla_4 = \overline{A}_{44}H_3 +\overline{A}_{65}H_5\mbox{ if }K\geq 3,
\nonumber
\\
&&
\nabla_5 = \overline{A}_{34}H_2+\overline{A}_{55}H_4 +\overline{A}_{76}H_6\mbox{ if }J\geq 3,
\nonumber
\\
&&
\nabla_6 = \overline{A}_{45}H_3+\overline{A}_{66}H_5 +\overline{A}_{87}H_7\mbox{ if }K\geq 4.
\nonumber
\end{eqnarray}
Also,
\begin{eqnarray*}
&&
\nabla_{re} =0\mbox{ if } r\leq 3,
\
\nabla_{4e} = \left[4^2\right]_0\ e\left(4^2\right),
\nonumber
\\
&&
\nabla_{5e} = [45]_0\ e(45) + [34]_1\ e(34),
\nonumber
\\
&&
\nabla_{6e} = \sum\left\{ [\pi]_0\ e(\pi):\ \pi=5^2,46,4^3\right\}
+\sum\left\{ [\pi]_1\ e(\pi):\ \pi=4^2,35\right\} + \left[3^2\right]_2\ e\left(3^2\right),
\nonumber
\end{eqnarray*}
where the  $e(\pi)$, $[\pi]_i$ needed for $e_r(x)$ are as follows:
Firstly, there is the special case
\begin{eqnarray}
h(ij \cdots) = H_{i+j+\cdots-1}\mbox{ so that }
h \left(1^{i_1}2^{i_2}\cdots\right) = H_{1i_1+2i_2+\cdots-1}
\nonumber
\end{eqnarray}
by (\ref{recur}), where $H_{k \cdot r}=D^r H_k$.
The other expressions needed for $e_r(x)$ are:
\begin{eqnarray*}
&&
\mbox{For } r = 4 :\  \left[4^2\right]_0 =\overline{A}_{43}^2/2!,
\\
&&
f \left(4^2\right) = H_7-H_1 H_3^2,
\
g\left(4^2\right) =H_7-2H_3H_4 +H_1 H_3^2.
\\
&&
\mbox{For } r = 5 :\  [45]_0=\overline{A}_{43}\overline{A}_{54},
\\
&&
f(45) = H_8-H_1H_3H_4,
\
g(45)= H_8-H_3H_5-H_4^2+H_1H_3H_4,
\\
&&
[34]_1 = \overline{A}_{33}\overline{A}_{43},
\
f(34) = H_6 - H_1 H_2H_3,
\
g(34)= H_6-H_2H_4-H_3^2 +H_1H_2H_3.
\\
&&
\mbox{For } r = 6:\  \left[5^2\right]_0 = \overline{A}_{54}^2/2!,
\\
&&
f\left(5^2\right) = H_9-H_1H_4^2,
\
g\left(5^2\right)= H_9 -2H_4H_5 +H_1H_4^2,
\\
&&
[46]_0  = \overline{A}_{43} \overline{A}_{65},
\\
&&
f(46) = H_9-H_1H_3H_5,
\
g(46) =H_9-H_3H_6 -H_4H_5+H_1H_3H_5,
\\
&&
\left[4^3\right]_0 = \overline{A}_{43}^3/3!,
\
f\left(4^3\right) =H_{11}-3H_1H_3H_7-H_2H_3^3+3H_1^2H_3^3,
\\
&&
g\left(4^3\right) = H_{11}-3H_3H_8-3H_4H_7+3H_1H_3H_7+3H_3^2H_5
\\
&&
+6H_3H_4^2-9H_1H_3^2H_4-H_2H_3^3+3H_1^2H_3^3,
\\
&&
\left[4^2\right]_1 = \overline{A}_{43}\overline{A}_{44},
\\
&&
[35]_1 = \overline{A}_{33}\overline{A}_{54},
\
f(35) = H_7-H_1 H_2 H_4,
\
g(35)= H_7- H_3 H_4 -H_2 H_5,
\\
&&
\left[3^2\right]_2 = \overline{A}_{33}^2/2!,
\
f\left(3^2\right) = H_5-H_1 H_2^2,
\
g\left(3^2\right) = H_5-2H_2H_3 +H_1 H_2^2.
\end{eqnarray*}
\end{theorem}

Now choose $w=1$, $\widehat{w}=G_m/m$, where $ G_m$ is a gamma variable with mean $m$.
By (\ref{1.22}),
\begin{eqnarray*}
&&
{A}_{r,r-1}^{JK} ={A}_{r,r-1,\theta}-(r-1)! \left( {A}_{32\theta}/2 \right)^{r-2},
\
r\geq 4,
\nonumber
\\
&&
{A}_{rr}^{JK} = {A}_{rr\theta}^{JK}={A}_{rr\theta}+d_{r1}{A}_{r,r-1,\theta},
\
r \geq 2,
K\geq 2,
\\
&&
{A}_{r,r+1}^{JK} = {A}_{r,r+1\theta}^{JK}
={A}_{r,r+1\theta}+d_{r1}{A}_{rr\theta}+d_{r2}{A}_{r,r-1,\theta},
\
r \geq 2,
K\geq 3,
\end{eqnarray*}
where
\begin{eqnarray}
d_{r1} = (-r/2){A}_{23\theta},
\
d_{r2} =(-r/2){A}_{23\theta} +{-r/2\choose 2} {A}_{22\theta}^2.
\nonumber
\end{eqnarray}
Sections 2 to 5 deal with the {\it general} case (\ref{now1})-(\ref{general}).
Section 2 gives simple
formulas for ${\bf H}$ and ${\bf a}$ of (\ref{notHe}) and (\ref{ar})
for $X$ a gamma variable with mean $m$,
and so for $X$ of (\ref{now}).
Examples in Section 2 include the distribution and quantiles of the sample variance
and the Studentized mean for non-normal populations.

Section 3 re-expresses $h_r(x,L)$   using a change of notation that does
away with the fractional coefficients in all of the papers referred to above.
Fractions are  eliminated by giving results, not in terms of $(l_1,l_2, \ldots)$, but in terms of
\begin{eqnarray}
\left[1^{i_1} 2^{i_2}\cdots\right] = \left(L_1^{i_1}/i_1!\right) \left(L_2^{i_2}/i_2!\right) \cdots,
\label{[]}
\end{eqnarray}
where $L_r =l_r/r!$.
We shall prove the following theorem.

\begin{theorem}
With notation as above,
\begin{eqnarray}
h_r(x,L) = \sum_{k=r,r+2, \ldots, 3r} C_{rk} H_{k-1}(x),
\label{hh}
\end{eqnarray}
where
\begin{eqnarray}
C_{rk}  =   \sum \left\{ [\pi]:\ i\in  {\cal H}_{rk} \right\},
\nonumber
\end{eqnarray}
and ${\cal H}_{rk}$ is the set of all partitions $\pi=1^{i_1}\cdots k^{i_k}$ of $k$ such that
\begin{eqnarray}
S(1) i_1+\cdots +S(k) i_k=r,
\nonumber
\end{eqnarray}
where
\begin{eqnarray}
S(r)= rI(r\leq 2) +(r-2)I(r\geq 3).
\nonumber
\end{eqnarray}
For example,
\begin{eqnarray}
C_{00} =1,
\
C_{rr} = \sum_{0\leq i\leq r/2} \left[1^{r-2i} 2^{i}\right].
\label{rr}
\end{eqnarray}
The other $C_{rk}$ are obtained from  these using
\begin{eqnarray}
&&
C_{r,r+2i} = \sum_{j=0}^{r-i}  C_{jj} {b}_{r-j, i} \left(\overline{L}\right),
\
1 \leq i \leq r,
\
\overline{L}_i=L_{i+2},
\
{b}_{ri}(x) = \widehat{B}_{ri}({x})/i!
\label{rr+2i}
\end{eqnarray}
and  $\widehat{B}_{ri}(x)$ is the ordinary Bell polynomial - see Appendix A.
\end{theorem}

Note that (\ref{rr}) and (\ref{rr+2i}) provide convenient ways  to calculate $h_r$ using  MAPLE, say.

Section 4 proves the following theorem.

\begin{theorem}
With notation as above, $f_r(x,L)$ and $g_r(x,L)$ of  (\ref{f})-(\ref{g}) can be expressed in the form
\begin{eqnarray}
e_r(x,L) = \sum \left\{ [\pi]\ e(\pi):\ \pi\in S_r(e)\right\},
\label{fgpi}
\end{eqnarray}
where  $S_r(e)$ is a set of partitions $\pi$ of $r, r+2, \ldots, 3r$.
The coefficients
$f(\pi)$, $g(\pi)$ are polynomials in ${\bf H}$.
\end{theorem}

Alternatively,  $ f(\pi)$, $g(\pi)$ can be written as polynomials in
${\bf a}$, as done in  Appendix D.
$f(\pi)$ are obtained via functions $c_r=c_{rX}(x)$ introduced by Hill and Davis (1968).
We express these in terms of  ${\bf H}$.

Section 5 gives  $f_r(x)$ and $g_r(x)$ of  (\ref{fff})-(\ref{ggg}).

Section 7 extends (\ref{hhh}) and (\ref{h}) to  expansions for the density and other
derivatives of $P_n(x)$.

Hill and Davis (1968) gave a different motivation: the distribution of the
likelihood ratio has an expansion of the form (\ref{hhh}),
with $X$ chi-square, or
equivalently, gamma.
Given an expansion of the form (\ref{hhh}),
they derive  (\ref{fff}) and (\ref{ggg}) giving $f_r$, $g_r$ in terms of $(h_1, h_2, \ldots)$.
Withers and Nadarajah (2012) simplified their results using Bell polynomials.

Another such example is when $X$ is {\it symmetric} about zero, such as $N$, or Student's $t$.
In this case $H_r$ is an even function for $r$ even, so that $h_r$
is an odd function for $r$ odd, and $\overline{Y}_n=|Y_{01\theta}|$ satisfies
\begin{eqnarray*}
\Pr \left( \overline{Y}_n\leq x \right) = \overline{P}(x)-\overline{p}(x)
\sum_{r=1}^\infty n^{-r}h_{2r}(x),
\end{eqnarray*}
where $\overline{P}(x)=\Pr(|X|\leq x) =P(x)-P(-x)$ has density $\overline{p}(x)=2{p}(x)$.

Appendix B gives the interesting expression for $H_r$ in terms of ${\bf a}$,
\begin{eqnarray*}
H_r(x)=(a_1-D)^r\ 1=(-1)^rB_r \left(-{\bf a}\right),
\
r\geq 0,
\end{eqnarray*}
where $B_r$ is the complete Bell polynomial,
and an inverse formula for  $a_r$ in terms of ${\bf H}$.
It also gives the derivatives of $H_r$ in terms of ${\bf H}$ using the functions
\begin{eqnarray}
b_{r}= \left(a_1+D\right)^r\ 1 = B_r({\bf a}),
\
r\geq 0.
\label{arH}
\end{eqnarray}
Appendix C gives $f(\pi)$ and  $g(\pi)$ needed in (\ref{fgpi}) in terms of ${\bf H}$.
Appendix D gives the same but in terms of ${\bf a}$.
Appendix E gives $e(\pi)$ for $e=f$, $g$ when $X$ is a gamma variable.
Appendix F specializes to $P(x)=\Phi(x) = \Pr(N\leq x)$, the standard normal distribution,
the choice used by Cornish and Fisher (1937) and
Fisher and Cornish (1960).
We give formulas for some  $ f(\pi)$, $g(\pi)$ that do not hold
when $X$ is non-normal.
The generating function is
$B_N(x,t)=e^{xt-t^2/2}=\mathbb{E}\ e^{(x+iN)t}$ so that $H_r=He_r$, the $r$th Hermite polynomial,
\begin{eqnarray}
He_r(x)=\phi(x)^{-1} (-D)^r \phi(x) =\mathbb{E}\ (x+iN)^r,
\label{He}
\end{eqnarray}
as noted by  Withers (2000).
Withers and Nadarajah (2011) also extended the results of Cornish and Fisher  (1937)
to general  $X$ and gave the recurrence relation
\begin{eqnarray}
H_r= J_1 H_{r-1},
\
r\geq 1,
\label{recur}
\end{eqnarray}
where $J_1=H_1-D$; its Sections 4 and 5 gave $H_r$ for $X$ a standardized gamma or a Student random variable,
using $\psi_{r+1}$ for $H_r$.
Its Appendix A gave $e_r(x,L)$ for $r\leq 4$, $e=h,f,g$.

We use the notation
\begin{eqnarray}
&&
[\alpha]_j = \Gamma(\alpha+1)/\Gamma(\alpha-j+1) =\alpha(\alpha-1)\cdots (\alpha-j+1),
\label{[]j}
\\
&&
(\alpha)_j  = \Gamma(\alpha+j-1)/\Gamma(\alpha-1) =\alpha(\alpha+1)\cdots (\alpha+j-1).
\nonumber
\end{eqnarray}
When changing variables, say from $Y$ to  $X=(Y-\mu)/\sigma$, it is convenient to set $P_X(x)= \Pr(X\leq x)$, $y=\mu+\sigma x$.
Then,
\begin{eqnarray}
&&
p_{X}(x) = \sigma p_Y(y),
\
a_{X}(x)=-\ln\sigma+a_Y(y),
\nonumber
\\
&&
A_{X}(x,t) = -\ln\sigma+ A_{Y}(y,\sigma t),
\
B_{X}(x,t)= B_{Y}(y,\sigma t),
\nonumber
\\
&&
H_{rX}(x) = \sigma^r H_{rY}(y),
\
a_{rX}(x)=\sigma^r a_{rY}(y),
\
c_{r+1,X}(x)=\sigma^r c_{rY}(y).
\label{dize}
\end{eqnarray}

\section{Expansions about the gamma, with examples}
\setcounter{equation}{0}

$Y_w$ needs to have a shape parameter if we want to reduce $A_{32}$ to zero.
Expansions about $X$ a standardized gamma or $\chi^2$ were given in
Section 4 of Withers and Nadarajah (2011).
However, it is easier to  evaluate first ${\bf a}$ and then ${\bf H}$.
Theorem 2.1 gives explicit formulas for these for gamma random variables.

\begin{theorem}
Let $G$ be a gamma random variable with mean $m$ and density $y^{m-1}e^{- y} /\Gamma (m)$ on $(0, \infty)$.
So,
\begin{eqnarray}
\kappa_r(G)= (r-1)!m,
\
A_{riw}=(r-1)! m^{1-r/2}
\label{hrg}
\end{eqnarray}
for $\widehat{w}=G/m$, $w=1$.
Set
\begin{eqnarray}
\alpha=m-1,
\
\overline{y}=-1/y.
\label{am}
\end{eqnarray}
For $X=G$, $a$, $a_r$ of  (\ref{ar}), $H_r$ and
the generating functions (\ref{agen}) and (\ref{Hgen}) are given by
\begin{eqnarray}
&&
a_G(y)  = y-\alpha\ln y+\ln \Gamma(\mu),
\
A_G(y,t)-a_G(y)=t-\alpha\ln \left(1-\overline{y}t\right),
\nonumber
\\
&&
a_{rG}(y) = I(r = 1) + (r-1)!\alpha \overline{y}^r,
\
r\geq 1,
\nonumber
\\
&&
B_G(y,t) = \left( 1+ \overline{y} t \right)^\alpha e^{t},
\nonumber
\\
&&
H_{rG}(y) = \sum_{j=0}^r {r\choose j}  [\alpha]_j \overline{y}^{j}
={}_2F_0 \left( -r,-\alpha,\overline{y} \right),
\label{gee}
\end{eqnarray}
where ${}_pF_q$ is the generalized hypergeometric distribution (see Section 9.14 of  Gradshteyn and Ryzhik (1965)).
Also
\begin{eqnarray}
Y_{JKw}=(G-\mu)/\sigma,
\label{41}
\end{eqnarray}
where $\sigma^2=m/s_{2Kw}$ and $\mu=m-s_{1Jw}(m/s_{2K})^{1/2}$.
By (\ref{dize}), for $X=Y_{JKw}$,
\begin{eqnarray*}
&&
H_r=H_{rX}(x)=\sigma^r H_{rG}(y),
\
a_r=\sigma^r a_{rG}(y) \mbox{ at } y=\mu+\sigma x,
\\
&&
c_{r+1,G} = r! \left(1+r\overline{y}\right) + \sum_{i=2}^{r} c_{r+1,i} \overline{y}^i,
\
r\geq 1.
\end{eqnarray*}
In particular,
\begin{eqnarray*}
&&
c_{r+1,r} = \prod_{k=1}^r (k\alpha+k-1),
\\
&&
c_{42} = \alpha(18\alpha+7),
\
c_{52} = \alpha(72\alpha+23),
\
c_{53} = 2\alpha(2\alpha+1)(24\alpha+11),
\\
&&
c_{62} = 2\alpha(600\alpha+163),
\
c_{63} = 2\alpha \left(489\alpha^2+600\alpha+101\right),
\\
&&
c_{64} = 3\alpha(2\alpha+1) \left(100\alpha^2+ 113\alpha+32\right).
\end{eqnarray*}
\end{theorem}

Matching skewness by (\ref{r3}) and  (\ref{hrg}) gives
\begin{eqnarray*}
A_{11}=A_{11\theta},
\
A_{22}=A_{22\theta},
\
A_{r,r-1}=A_{r,r-1,\theta}-\tau^{1-r/2}(r-1)!
\end{eqnarray*}
for $r\geq 3$.
So, $A_{32}=0$ if we choose $m=n\tau$ with $\tau^{1/2}=2/A_{32\theta}$, in (\ref{41}).
Then,
\begin{eqnarray*}
l_{0r} =A_{r,r-1}=A_{r,r-1,Y}-(r-1)! \left(A_{32Y}/2\right)^{r-2}
\end{eqnarray*}
for $r\geq 4$.
In particular, $A_{43} =A_{43Y}-3A_{32Y}^2/2$, $A_{54}=A_{54Y}-3A_{32Y}^3$, and $A_{65}=A_{65Y}-15A_{32Y}^4/2$.

\begin{example}
The $F$ variable is defined by
$F_{n_1,n_2}=(\chi^2_{n_1}/n_1)/(\chi^2_{n_2}/n_2)$,
where the chi-square variable $\chi^2_{n_1}$ is independent of  $\chi^2_{n_2}$.
Wishart (1947) gave expansions in powers of $n_1^{-1}$, $n_2^{-1}$ for the
cumulants of $Z=2^{-1}\ln F_{n_1,n_2}$.
Setting $n=n_1+n_2$ say, and $f_i=n/n_i$, it follows that $\widehat{\theta}=Z$
satisfies (\ref{stand}) holds with $a_{10\theta}=\theta=0$ and the non-zero $a_{ri\theta}$ given by
\begin{eqnarray}
&&
a_{rr\theta} = \left[f_2^{r}+(-1)^rf_1^{r}\right](r-1)!/2,
\nonumber
\\
&&
a_{r,2j+r-1,\theta} = 2  \left[f_2^{2j+r-1}+(-1)^rf_1^{2j+r-1}\right] (-4)^{j-1}B_j(2j+r-2)!/(2j)!,
\
(j,r)\neq (0,1).
\nonumber
\end{eqnarray}
The $B_j$ are given by $B_0=-1$, $B_1=1/6$, $B_2=1/30$, $B_3=1/42$, $B_4=1/30$, $B_5=5/66$.
(This is not the current notation for the Bernoulli numbers.
Apart from $B_0$,  his $B_j$ is what we call $|B_{2j}|$ today.)
So, $a_{21\theta}=(f_2+f_1)/2$.
Now redefine $n$ by $n=2/\sum_{i=1}^2 n_i^{-1}$, the harmonic mean of $n_1$, $n_2$.
(We can do this as $n$ is arbitrary provided that
it has magnitude $\min(n_1,n_2)$.)
So, now $a_{21\theta}=1$.
Writing $a_{ri}=a_{ri\theta}$, the coefficients needed for $e_r(x)$ are
\begin{small}
\begin{eqnarray*}
&&
e_1: \ a_{11 }= \left(f_2-f_1\right)/2, a_{32 }= \left(f_2^2-f_1^2\right)/2,
\\
&&
e_2: \ a_{22 }= \left(f_2^2+f_1^2\right)/2, a_{43}=f_2^3+f_1^3,
\\
&&
e_3: \ a_{12 } = \left(f_2^2-f_1^2\right)/6, a_{33 }= f_2^3-f_1^3, a_{54 }=3\left(f_2^4-f_1^4\right),
\\
&&
e_4: \ a_{23 }=\left(f_2+f_1\right)/3, a_{44 }=3\left(f_2^4+f_1^4\right), a_{65 }=12\left(f_2^5+f_1^5\right),
\\
&&
e_5: \ a_{13 }=0, a_{34 }=f_2^4-f_1^4, a_{55 }=12\left(f_2^5-f_1^5\right), a_{76 }=60\left(f_2^6-f_1^6\right),
\\
&&
e_6: \ a_{24}=0, a_{45 }=4\left(f_2^5+f_1^5\right), a_{66}=60\left(f_2^6+f_1^6\right),
a_{87}=360\left(f_2^7+f_1^7\right).
\end{eqnarray*}
\end{small}
Cornish and Fisher (1937, page 319) and Fisher and Cornish (1960, page 216)
denote $a_{21\theta}/n$ by $\sigma/2$.
They illustrated the quantile expansion (\ref{quan}) for $n_1=24$, $n_2=60$, $P(x)=0.95$,
giving columns 1 to 4 of Table 2.1 using $P=\Phi$,
the normal distribution.
(They give the exact value as $.2653 4844\cdots$.)

The picture is less rosy if the degrees of freedom are small.
In this case the series must be truncated when divergence begins, giving
an upper bound to the accuracy achievable, as illustrated by Table 2.2 for degrees of freedom 5 and 5.

We can write $\chi^2_n/n=G_m/m$, where $G_m$ is a gamma variable with mean $m=2n$.
So, switching to $\gamma_i=2f_i=n/m_i$ and
$\widehat{\theta}=\ln (G_{m_1}/m_1)-\ln (G_{m_2}/m_2)$, where  $G_{m_1}$ is independent of $G_{m_2}$,
it follows that  (\ref{stand}) holds with $a_{10\theta}=\theta=0$, and
the non-zero $a_{ri\theta}$ given by
\begin{eqnarray}
&&
a_{rr\theta} = \left[\gamma_2^{r}+(-1)^r\gamma_1^{r}\right](r-1)!/2,
\nonumber
\\
&&
a_{r,2j+r-1,\theta} = (-1)^{j-1} \left[\gamma_2^{2j+r-1}+(-1)^r\gamma_1^{2j+r-1}\right]B_j (2j+r-2)!/(2j)!,
\
(j,r)\neq (0,1).
\nonumber
\end{eqnarray}
For example, the leading coefficients are
\begin{eqnarray*}
a_{r,r-1,\theta} =(r-2)! \left[\gamma_2^{r-1}+(-1)^r\gamma_1^{r-1}\right],
\
r\geq 2.
\end{eqnarray*}
So, $a_{21\theta}=\gamma_2+\gamma_1=1$ if we  redefine  $n$ as $n=1/\sum_{i=1^2} m_i^{-1}=m_1m_2/(m_1+m_2)$,
so that now   (\ref{riY}) holds with $ A_{ri\theta}= a_{ri\theta}$.
($\gamma_i$ is still given by $n/m_i$.)
\end{example}

\noindent
[Tables 2.1 and 2.2 about here.]

We end with two non-parametric examples.
Suppose that we have a random sample $X_1, \ldots, X_n$
of size $n$
from an unknown distribution
$F$ with mean $\mu=\mu(F)$
and finite central moments $\mu_r=\mu_r(F)$.
Their empirical estimates are
\begin{eqnarray*}
\mu\left(F_n\right)=\overline{X}=n^{-1}\sum_{j=1}^n X_j,
\
\mu_r\left(F_n\right)=n^{-1}\sum_{j=1}^n \left(X_j-\overline{X}\right)^r,
\end{eqnarray*}
where $F_n$ is the  empirical distribution.
Set $\nu_r=\mu_r/\mu_2^{r/2}$.
By Withers (1983), for $T(F)$ a smooth functional,
the cumulants of  $\widehat{\theta}=T(F_n)$
have an expansion of the form (\ref{stand}) with  ${\theta}=T(F)$
and the leading cumulant coefficients $a_{ri\theta}$ given by Theorem 3.1 there.

\begin{example}
The distribution and quantiles
of the sample variance, $\widehat{\theta}={\mu}_2(F_n)$.
(After scaling, this is equivalent to choosing the unbiased estimate
$s^2=n \mu_2(F_n)/(n-1)$.)
So, $\theta=\mu_2$, $a_{21\theta}=\mu_4-\mu_2^2$.
The leading $a_{ri} = a_{ri\theta}$ of (\ref{riY})
are given by Example 4.2 of Withers (1983) in terms of $\delta=\nu_4-1$:
\begin{small}
\begin{eqnarray*}
&&
e_1: \ a_{11 }= -\mu_2,
\
a_{32 }= \mu_6-3\mu_4\mu_2+2\mu_2^3-6\mu_3^2,
\\
&&
e_2: \ a_{22 }= 4\mu_2^2-2\mu_4,
a_{43} =  \mu_8-4\mu_6\mu_2+12\mu_4\mu_2^2-3\mu_4^2-24\mu_5\mu_3+96\mu_3^2\mu_2 - 6\mu_2^4,
\\
&&
e_3: \ a_{12 }=0,
\
a_{33 }=-3 \mu_6 +21\mu_4\mu_2-26\mu_2^3 +18\mu_3^2,
\
a_{54 }=\mu_{10}- 5\mu_8\mu_2  -40\mu_7\mu_3
\\
&&
-10\mu_6\mu_4 +20\mu_6 \mu_2^2 - 30\mu_5^2 +480\mu_5\mu_3  +360\mu_4\mu_3^2 +
30\mu_4^2 -60\mu_4\mu_2^3 -1560\mu_3^2\mu_2^2 +24\mu_2^5.
\end{eqnarray*}
\end{small}
\end{example}

\begin{example}
The distribution and quantiles of the Studentized mean,
$Y_{01\theta}= n^{1/2} \widehat{\theta}$,
where $\widehat{\theta}=\mu_2(F_n))^{-1/2}(\overline{X}-\mu)=T_0(F_n)$ say.
(For a normal sample, $(1-n^{-1})^{1/2} Y_{01\theta} \sim t_{n-1}$,
but otherwise it is simpler to calculate  $A_{ri\theta}$ for $Y_{01\theta}$.)
So, $ a_{10\theta}=0$, $a_{21\theta}=1$, and the other leading
$a_{ri}=a_{ri\theta}=A_{ri\theta}$ of (\ref{riY})
are given in Example 1.2 of Withers (1989b):
\begin{small}
\begin{eqnarray*}
&&
e_1: \ a_{11 }=-\nu_3/2,
\
a_{32 }=-2\nu_3,
\\
&&
e_2: \ a_{22 }= 3+7\nu_3^2/2,
\
a_{43}=12-2\nu_4+12\nu_3^2,
\\
&&
e_3: \ a_{12 }= \left(-25\nu_{3}+6\nu_{5}-15\nu_{3}\nu_{4}\right)/16.
\end{eqnarray*}
\end{small}
(This reference also gives the leading $h_r$, $f_r$, $g_r$ when $X=N$.)
\end{example}

\section{$h_r(x,L)$ in terms of sums of partitions}
\setcounter{equation}{0}

This section can be skipped on a first reading.
It proves Theorem 1.3.

\noindent
{\bf Proof of Theorem 1.3}
Formulas for ${\cal H}_{rk}$ are easily derived from (\ref{rr}) and (\ref{rr+2i}).
Here, we only note that
each ${\cal H}_{rk}$ is a distinct set of partitions of $k$, and
\begin{eqnarray*}
\ \cup_r\ {\cal H}_{rk} = \ \cup_{0\leq i\leq k/2}\ {\cal H}_{k-2i,k}
\end{eqnarray*}
is the set of all the partitions of $k$.

When $X=N$, equation (3.1) of Withers (1984) gave the formula
\begin{eqnarray}
h_r(x,L) = \sum L_{r_1}\cdots L_{r_j} \ H_{r_1+\cdots+ r_j-1}(x)/j!
\label{hL}
\end{eqnarray}
summed over $j\geq 1$, $r_1\geq 1$, $\ldots$, $r_j\geq 1$, $S(r_1)+\cdots +S(r_j)=r$.
Its proof follows from the Charlier differential series - see  Withers and Nadarajah (2012),
so it remains valid for general $X$.

Let us rewrite (\ref{hL}) in the form
\begin{eqnarray}
h_r(x,L)  = \sum_{k=1}^{3r} C_{rk} H_{k-1}(x),
\
C_{rk}  = \sum_{j=1}^r C_{rkj},
\label{hL2}
\end{eqnarray}
where
\begin{eqnarray}
C_{rkj} = \sum L_{r_1}\cdots L_{r_j}/j!
\nonumber
\end{eqnarray}
is summed over $r_1\geq 1, \ldots, r_j\geq 1, S(r_1)+\cdots +S(r_j)=r$.

Now suppose that $\{r_1, \ldots, r_j\}$ consists of $i_1\ 1s$, $i_2\ 2s$, $\cdots$, $i_k\ ks$.
The number of ways this can arise is the multinomial
coefficient, ${j\choose i_1\cdots i_k}$.
So, we can rewrite $ C_{rkj}$  using the $[\ ]$ notation of (\ref{[]}) in the form
\begin{eqnarray}
C_{rkj}  = \sum \left[ 1^{i_1}\cdots k^{i_k} \right]
\nonumber
\end{eqnarray}
summed over
\begin{eqnarray*}
i_1+\cdots +i_k=j,
\
1 i_1+\cdots +k i_k=k,
\
S(1) i_1+\cdots +S(k) i_k=r.
\end{eqnarray*}
(For $\pi$ a partition, Hill and Davis denote $[\pi]$ by $l_\pi$.)
The last constraint can be written $ i_1+2i_2 + \sum_{a=3}^k (a-2) i_a=r$.
So, these three constraints can be written
\begin{eqnarray*}
i_3+\cdots +i_k=j-i_1-i_2,
\
1 i_1+\cdots +k i_k=k,
\
i_1+2i_2 + \sum_{a=3}^k (a-2)i_a=r.
\end{eqnarray*}
So,
\begin{eqnarray}
C_{rk} = \sum_{i_1+2i_2\leq r}  \left[1^{i_1} 2^{i_2}\right]\ C_{rki_1i_2},
\label{Crk}
\end{eqnarray}
where
\begin{eqnarray*}
C_{rki_1i_2}   = \sum \ \left[3^{i_3}\cdots k^{i_k}\right]
\end{eqnarray*}
is summed over
\begin{eqnarray}
&&
\sum_{a=3}^k i_a=(k-r)/2=K\mbox{ say, }
\nonumber
\\
&&
\sum_{a=3}^k (a-2)i_a=r-i_1-2i_2=R\mbox{ say}.
\nonumber
\end{eqnarray}
By (\ref{B}), $C_{rki_1i_2} = {b}_{RK}(\overline{L})$  at $\overline{L}_i  =L_{i+2}$.

By (\ref{rr}),
\begin{eqnarray*}
&&
C_{11}  = [1],
\
C_{22}  = \left[1^2\right]+ [2],
\
C_{33}  = \left[1^3\right]+ [12],
\
C_{44}  = \left[1^4\right]+ \left[1^2 2\right]+ \left[2^2\right],
\\
&&
C_{55} = \left[1^5\right]+ \left[1^3 2\right] + \left[1 2^2\right],
\
C_{66}  = \left[1^6\right] + \left[1^4 2\right] + \left[1^2 2^2\right] + \left[2^3\right],
\end{eqnarray*}
and we can write (\ref{Crk}) as
\begin{eqnarray}
C_{rk} =  \sum_{j=0}^r  C_{jj} {b}_{r-j, K} \left(\overline{L}\right),
\nonumber
\end{eqnarray}
giving (\ref{rr+2i}),  since $\widehat{B}_{RK}=0$ for $R<K$, and $C_{rk}=0$ if $k-r$ is odd or $k<r$.
So,
\begin{eqnarray}
h_r(x,L)
&=&
\sum_{k=r,r+2, \ldots, 3r} C_{rk} H_{k-1}(x)
\nonumber
\\
&=&
\sum_{i=0}^r \left[ C_{rk} H_{k-1}(x)\right]_{k=r+2i}   =
\sum_{j=0}^r \left[ C_{rk} H_{k-1}(x) \right]_{k=3r-2j}.
\nonumber
\end{eqnarray}
The proof is complete.
\
$\Box$

We can rewrite  (\ref{rr+2i}) as
\begin{eqnarray}
C_{r,3r-2i} = \sum_{j=0}^{i} C_{jj} {b}_{r-j, r-i},
\
r\geq i\geq 0.
\label{r3r-2i}
\end{eqnarray}
Using Appendix A, we obtain some special cases of the results in Theorem 1.3.

\begin{corollary}
With notation as above,
\begin{eqnarray}
&&
C_{r,r+2} = \sum_{j=0}^{r-1} C_{jj} \alpha_{r-j},
\label{r,r+2}
\\
&&
C_{r,3r} = b_{rr} \left(\overline{L}\right) = \left[3^r\right],
\label{r,3r}
\\
&&
C_{r,3r-2} =  \left[1 3^{r-1}\right] + b_{r,r-1}\left(\overline{L}\right),
\
b_{r,r-1}\left(\overline{L}\right) = \left[3^{r-2} 4\right],
\label{r,3r-2}
\\
&&
C_{r,3r-4} =  C_{22} \left[3^{r-2}\right] + \left[1 3^{r-3} 4\right] +
b_{r,r-2} \left(\overline{L}\right),
\
b_{r,r-2}\left(\overline{L}\right) = \left[3^{r-4} 4^2\right] + \left[3^{r-3} 5\right],
\label{r,3r-4}
\end{eqnarray}
where $\alpha_r= {b}_{r1}(\overline{L})=L_{r+2}$ for $r\geq 1$, and a term with a negative
power of three is discarded.
For example,
\begin{eqnarray*}
&&
C_{24} = L_4+L_1L_3=[4]+[13],
\
C_{26}=\left[3^2\right],
\\
&&
C_{46} =L_6+L_1L_5+C_{22}L_4+C_{33}L_3 =[6]+[15]+\left[1^24\right]+[24]+\left[1^33\right]+[123],
\\
&&
C_{48} = \left[4^2\right]+[35]+[134]+C_{22}\left[3^2\right],
\
C_{22}\left[3^2\right]=\left[1^23^2\right]+\left[23^2\right],
\\
&&
C_{5,11} = C_{22} \left[3^3\right]+ C_{11}\left[3^2 4\right]+\left[3 4^2\right]+\left[3^2 5\right]
 =  \left[1^2 3^3\right]+\left[1 3^2 4\right] + \left[2 3^3\right]+\left[3 4^2\right]+\left[3^2 5\right].
\end{eqnarray*}
So, using the $[\cdot]$ notation, {\it all the numerical coefficients of the components of $h_r(x,L)$ are 1}:
\begin{eqnarray*}
&&
h_1(x,L) = [1]+[3]H_2,
\\
&&
h_2(x,L) = \left(\left[1^2\right]+[2]\right)H_1 + ([13]+[4])H_3 + \left[3^2\right]H_5,
\\
&&
h_3(x,L) = \left(\left[1^3\right]+[12]\right)H_2 +
\left(\left[1^2 3\right]+[23] +[14] +[5]\right)H_4 +
\left(\left[1 3^2\right] +[34]\right) H_6
\\
&&
+ \left[3^3\right] H_8,
\\
&&
h_4(x,L) = \left(\left[1^4\right]+\left[1^2 2\right] +\left[2^2\right]\right) H_3+
\left(\left[1^3 3\right] + \left[1^2 4\right]+[123] +[15] +[24] +[6]\right) H_5
\\
&&
+ \left( \left[1^2 3^2\right] +[134] + \left[2 3^2\right] +[35] + \left[4^2\right]\right) H_7 +
\left(\left[1 3^3\right]+\left[3^2 4\right]\right) H_9 + \left[3^4\right] H_{11},
\end{eqnarray*}
and so on.
\end{corollary}

However, it is safer to calculate $H_r(x,L)$ in MAPLE using
(\ref{hh}), (\ref{rr}), (\ref{rr+2i}), (\ref{r3r-2i})-(\ref{r,3r-4}) to avoid the chance of missing a term.

Note how each $C_{rk}$ sums over a distinct set of partitions of $k$,
and how  $\sum_r C_{rk}$  sums over all distinct set of partitions of $k$.
When $X=N$, the 2nd term in $h_r(x,L)$, $C_{r,r+2}H_{r+1}(x)$,
also occurs in $f_r(x,L)$ and  $g_r(x,L)$.

To convert a term to the form given by
Cornish and Fisher (1937),  Fisher and Cornish (1960) and Withers (1984),
it is only necessary to substitute
\begin{eqnarray*}
\left[1^{i_1} 2^{i_2}\cdots\right] = \prod_{k=1} \left(l_k/k!\right)^{i_k}/i_k!
= \left(1^{i_1} 2^{i_2}\cdots\right) \prod_{k=1} l_k^{i_k}/r!,
\end{eqnarray*}
where $r =\sum ki_k$  and $(1^{i_1} 2^{i_2}\cdots) =r!/ \prod_{k=1} k!^{i_k}i_k!$
is the partition function.

\section{Expressions for $f_r(x,L)$  and $g_r(x,L)$  in terms of $(L, {\bf H})$}
\setcounter{equation}{0}

We begin by giving the proof of Theorem 1.4.

\noindent
{\bf Proof of Theorem 1.4}
By equation (5.6)  of  Withers and Nadarajah (2012),
\begin{eqnarray}
&&
f_r(x,L) =\sum_{k=1}^r (-1)^{k-1} c_k\ {b}_{rk}(h),
\label{ff}
\\
&&
g_r(x,L) =\sum_{k=1}^r (-1)^{k-1} D_k\ {b}_{rk}(h)
\label{gg}
\end{eqnarray}
for $b_{rk}$ of (\ref{rr+2i}), where $h_r=h_r(x,L)$.
So, (\ref{hh}), (\ref{ff}) and (\ref{gg}) can be written in the form (\ref{fgpi}),
where  $S_r(h)=\cup_k {\cal H}_{rk}$ is the set of partitions
$\pi=(i_1i_2\cdots)$ such that $S(i_1)+S(i_2)+\cdots=r$, $h(\pi)=H_{i_1+i_2+\cdots-1}$.
If $\pi$ is a partition of $k$, then
the coefficients $ f(\pi)$, $g(\pi)$ are
polynomials in $\{H_1, \ldots, H_{k-1}\}$.
The exception is $h(1) = f(1) =g(1)=H_0=1$.
\
$\Box$

Corollary 4.1 gives some particular cases of (\ref{ff})-(\ref{gg}) and $f(\cdot)$, $g (\cdot)$, $h (\cdot)$.

\begin{corollary}
With notation as above,
\begin{eqnarray}
&&
g_1(x,L) = f_1(x,L) =h_1(x,L) = [1]+[3]H_2,
\label{1}
\\
&&
f_4 = h_4-c_2 \left(h_1h_3+h_2^2/2\right) +c_3h_1^2h_2/2-c_4h_1^4/4!,
\nonumber
\\
&&
g_4 = h_4-D_2 \left(h_1h_3+h_2^2/2\right) +D_3h_1^2h_2/2-D_4h_1^4/4!.
\nonumber
\end{eqnarray}
Here, $c_k$ and $D_k$ are the function and operator
introduced by Hill and Davis (1968):
\begin{eqnarray}
&&
c_1 =1,
\
c_{k+1}=K_k\ c_k=K_k\cdots K_1 \ 1,
\nonumber
\\
&&
D_1 =1,
\
D_{k+1}=D_k\ J_k =J_1\cdots J_k,
\nonumber
\end{eqnarray}
where $K_k=kH_1+D$ and $J_k=kH_1-D$.
Using the expressions for $c_k$ in terms of ${\bf a}$ given in Withers and Nadarajah (2012), we obtain
\begin{eqnarray*}
&&
c_2 =H_1,
\
c_3=  3H_1^2-H_2,
\
c_4 =  15H_1^3-10H_1H_2+H_3,
\nonumber
\\
&&
c_5 = 105H_1^4-105H_1^2H_2 +15 H_1H_3+10H_2^2-H_4,
\nonumber
\\
&&
c_6 =945 H_1^5 -  1260 H_1^3H_2 + 210 H_1^2H_3 +280 H_1H_2^2 -35 H_2H_3 -21 H_1H_4 + H_5.
\nonumber
\end{eqnarray*}
The coefficient of $H_1^r$ in $c_{r+1}$ is $\mathbb{E}\ N^{2r}$.
That of $H_1^{r-2}H_2$ is $-(r-1)\mathbb{E}\ N^{2r}/3$.

Other particular cases are
\begin{eqnarray}
&&
h(k) =f(k) =g(k)=H_{k-1},
\
f\left(1^i\right)=0\mbox{ for }i\geq 2,
\label{hfg}
\\
&&
f(1,k+1) = H_k-H_1H_{k-1}=-H_{k \cdot 1},
\
g(1i) =0,
\label{1k+1}
\\
&&
f\left(1^{i-1}2\right) =\kappa_i({\bf H})=\sum_{j=1}^i (-1)^j (j-1)! B_{ij}({\bf H}),
\
i\geq 1.
\label{kapm}
\end{eqnarray}
\end{corollary}

The last formula, (\ref{kapm}), is just the familiar formula for $\kappa_i$ in terms of the
non-central moments, ${\bf m}=(m_1, m_2, \ldots)$:
\begin{eqnarray*}
&&
\kappa_1 = m_1,
\
\kappa_2 = m_2-m_1^2,
\
\kappa_3 = m_3- 3m_2m_1+2m_1^3,
\\
&&
\kappa_4 = m_4- 4m_3m_1-3m_2^2+12m_2m_1^2-6m_1^4,
\cdots
\end{eqnarray*}
as given by equation (3.42) of  Stuart and Ord (1987) up to $i=10$.
The formula
\begin{eqnarray*}
\kappa_i=\sum_{j=1}^i (-1)^j (j-1)! B_{ij}({\bf m}),
\end{eqnarray*}
and its inverse formula, $m_i=B_i({\bf \kappa})$, where ${\bf \kappa}=(\kappa_1, \kappa_2, \ldots)$,
were given by  Withers and Nadarajah (2009),
and in equation (2) of Comtet (1974, page 160).
For example, $f(1^22)=\kappa_3(H)=H_3-3H_1H_2+2H_1^3$.
By (\ref{1}), $h_1(x,L)=f_1(x,L)=g_1(x,L)$ is given by
\begin{eqnarray}
S_1(f) = S_1(g)= \{1,3\},
\
f(1)=H_{0}=1,
\
f(3)=H_{2}.
\nonumber
\end{eqnarray}
For $e(\pi)$, $S_r(e)$ needed by (\ref{fgpi})
to compute $e_r(x,L)$, $2\leq r\leq 6$, $e=f,g$, see Appendix C.

\section{Expansions for standardized estimates}
\setcounter{equation}{0}

Theorem 5.1 gives tools for calculating $f_r(x)$ and $g_r(x)$ of  (\ref{fff})-(\ref{ggg}).

\begin{theorem}
Expand $[\pi]$, $C_{rk}$, $e_r(x,L)$ in the form
\begin{eqnarray*}
[\pi]=\sum_{i=0}^\infty [\pi]_i n^{-i},
\
C_{rk}=\sum_{i=0}^\infty C_{rki} n^{-i},
\
e_r(x,L) = \sum_{i=0}^\infty e_{ri}(x) n^{-i}.
\end{eqnarray*}
Then, in terms of $\overline{A}_{ri}$ of (\ref{bar}), $l_{0r}$ of (\ref{ll}), and
$L_{0r} =l_{0r}/r!$, $[\pi](L)=[\pi]$, $C_{rk}(L)= C_{rk}$, we have for $e=h,f,g$,
\begin{eqnarray}
&&
[\pi]_0 = [\pi] \left(L_0\right),
\
C_{rk0}= C_{rk}\left(L_0\right),
\nonumber
\\
&&
[r]  =L_{r}=l_{r}/r!,
\nonumber
\\
&&
[r]_i = \overline{A}_{r,r+i-\delta},
\
\delta=I(r\geq 3),
\nonumber
\\
&&
e_{1i} = [1]_i +[3]_i H_2 =A_{1,1+i} +\overline{A}_{3,2+i}H_2,
\nonumber
\\
&&
h_{ri} = \sum_{k=r,r+2, \ldots, 3r} C_{rki}H_{k-1}.
\nonumber
\end{eqnarray}
To find $h_{r0}$, note that
when $l_1=l_2=0$, $C_{jj}=0$ for $j>0$, so that by (\ref{rr+2i}),
$C_{r,r+2i}= {b}_{ri}(\overline{L})$, giving
\begin{eqnarray*}
h_{r0} =h_r\left(x,L_0\right),
\end{eqnarray*}
where
\begin{eqnarray*}
h_r(x,L) = \sum_{i=1}^r  {b}_{ri} \left(\overline{L}\right) H_{r+2i-1}.
\end{eqnarray*}

Also,
\begin{eqnarray}
&&
\Delta_{1e} =\Delta_{2e}=0,
\
\Delta_{3e}  = h_{11}=A_{12}+A_{33}H_{2}/6,
\nonumber
\\
&&
\Delta_{4e} = e_{21},
\
\Delta_{5e} =  e_{12} +e_{31},
\
\Delta_{6e} = e_{22} +e_{41}.
\label{456e}
\end{eqnarray}
For example, for $i=1,2$ and $r\geq 3$,
\begin{eqnarray}
[ir]_1 = \overline{A}_{ii}\overline{A}_{rr}+\overline{A}_{i,i+1}\overline{A}_{r,r-1}
\nonumber
\end{eqnarray}
and
\begin{eqnarray*}
&&
\left[23^2\right]_1 = \overline{A}_{22}\overline{A}_{32} \overline{A}_{33}+ \overline{A}_{23} \overline{A}_{32}^2/2,
\\
&&
\left[2^2\right]_1  = \overline{A}_{22} \overline{A}_{23},
\\
&&
\left[1^2 4\right]_1 =  A_{11}^2 \overline{A}_{44}/2! +A_{11} A_{12} \overline{A}_{44},
\
\left[1^2 2\right]_1 =  A_{11}^2\overline{A}_{23}/2 +A_{11} A_{12} \overline{A}_{22},
\\
&&
[123]_1 =  A_{11} \overline{A}_{22}\overline{A}_{33} + A_{11} \overline{A}_{23}\overline{A}_{32}
+A_{12} \overline{A}_{22} \overline{A}_{32},
\\
&&
\left[1^2 3^2\right]_1 = A_{11}^2 \overline{A}_{32}\overline{A}_{33}/2 +A_{11} A_{12} \overline{A}_{32}^2/2,
\
\left[1^2 2\right]_1 = A_{11}^2 \overline{A}_{23}/2 +A_{11} A_{12} \overline{A}_{22},
\\
&&
\left[1^33\right]_1 = A_{11}^3\overline{A}_{33}/3! +A_{11}^2 A_{12} \overline{A}_{32}/2.
\end{eqnarray*}
The terms needed for $\Delta_{rh}$ for $r=4,5,6$ are
\begin{small}
\begin{eqnarray}
&&
r =4:\
h_{21} = \sum_{k=2,4,6} C_{2k1}H_{k-1},
\nonumber
\\
&&
C_{221} =  \left[1^2\right]_1 +\overline{A}_{23},
\
\left[1^2\right]_1  = A_{11} A_{12},
\nonumber
\\
&&
C_{241} = [4]_1 + [13]_1,
\
[4]_1 =\overline{A}_{44},
\
[13]_1 = A_{11} \overline{A}_{33}+ A_{12} \overline{A}_{32},
\nonumber
\\
&&
C_{261} =  \left[3^2\right]_1 = \overline{A}_{32} \overline{A}_{33}.
\nonumber
\\
&&
r = 5:\
e_{12} =A_{13} +\overline{A}_{34}H_2\mbox{ for }e=h,f,g,
\label{e12}
\\
&&
h_{31} =\sum_{k=3,5,7,9} C_{3k1} H_{k-1},
\nonumber
\\
&&
C_{331} = \left[1^3\right]_1  + [12]_1,
\
\left[1^3\right]_1 = A_{11}^2 \overline{A}_{12},
\
[12]_1 = A_{11} \overline{A}_{23} + {A}_{12} \overline{A}_{22},
\nonumber
\\
&&
C_{351} =  [5]_1 + [14]_1 + \left[1^2 3\right]_1 + [23]_1,
\
[5]_1 =\overline{A}_{55},
\nonumber
\\
&&
[14]_1 =  A_{11} \overline{A}_{44}+ A_{12}  \overline{A}_{43},
\nonumber
\\
&&
\left[1^2 3\right]_1 =  A_{11}^2 \overline{A}_{33}/2 +A_{11} A_{12} \overline{A}_{32},
\
[23]_1 = \overline{A}_{22} \overline{A}_{33}+ \overline{A}_{23} \overline{A}_{32},
\nonumber
\\
&&
C_{371} = \left[1 3^2\right]_1 +[34]_1,
\
\left[1 3^2\right]_1 =A_{11}\overline{A}_{32} \overline{A}_{33}+ A_{12}  \overline{A}_{32}^2/2,
\
[34]_1 =  \overline{A}_{32} \overline{A}_{44}+ \overline{A}_{33}  \overline{A}_{43},
\nonumber
\\
&&
C_{391} = \left[3^3\right]_1 =  \overline{A}_{32}^2 \overline{A}_{33}/2.
\nonumber
\\
&&
r =6:\
h_{22} =\sum_{k=2,4,6} C_{2k2}H_{k-1},
\nonumber
\\
&&
C_{222} = \left[1^2\right]_2+[2]_2,
\
\left[1^2\right]_2=A_{12}^2/2+ A_{11} A_{13},
\
\left[1^2\right]_2= \overline{A}_{24},
\nonumber
\\
&&
C_{242} = [4]_2 +[13]_2,
\
[4]_2=\overline{A}_{45},
\
[13]_2= A_{11} \overline{A}_{34}+ A_{12} \overline{A}_{33}+ A_{13} \overline{A}_{32},
\nonumber
\\
&&
C_{262} = \left[3^3\right]_2= \overline{A}_{32} \overline{A}_{34}+ \overline{A}_{33}^2/2,
\nonumber
\\
&&
h_{41} =  \sum_{k=4,6,8,10,12} C_{4k1}H_{k-1},
\nonumber
\\
&&
C_{441} = \left[1^4\right]_1 + \left[1^22\right]_1 + \left[2^2\right]_1,
\
\left[1^4\right]_1=A_{11}^3  A_{12}/3!,
\nonumber
\\
&&
\left[1^22\right]_1 =A_{11} A_{12} \overline{A}_{22}+ A_{11}^2 \overline{A}_{23}/2,
\
\left[2^2\right]_1 = \overline{A}_{22} \overline{A}_{23},
\nonumber
\\
&&
C_{461} = [6]_1 + [15]_1 + \left[C_{22}L_4\right]_1 + \left[C_{33}L_3\right]_1,
\
[6]_1=\overline{A}_{66},
\
[15]_1 =A_{11} \overline{A}_{55}+ A_{12} \overline{A}_{54},
\nonumber
\\
&&
\left[C_{22}L_4\right]_1 = A_{43} C_{221} + [4]_1 C_{220},
\
[4]_1=\overline{A}_{44},
\nonumber
\\
&&
\left[C_{33}L_3\right]_1 = \overline{A}_{32} C_{331} + \overline{A}_{33} C_{330},
\\
&&
C_{220} = A_{11}^2/2!+\overline{A}_{22},
\
C_{330} = A_{11}^3/3!+ A_{11} \overline{A}_{22},
\nonumber
\\
&&
C_{481} = C_{220} \left[3^2\right]_1 +C_{221} \left[3^2\right]_0 + [134]_1 + \left[4^2\right]_1 +  [35]_1,
\nonumber
\\
&&
\left[3^2\right]_0 =\overline{A}_{32}^2/2!,
\
[134]_1  =  A_{11} \overline{A}_{32}\overline{A}_{44} + A_{11} \overline{A}_{33}\overline{A}_{43}
+A_{12} \overline{A}_{32} \overline{A}_{43},
\nonumber
\\
&&
\left[4^2\right]_1 =  \overline{A}_{43} \overline{A}_{44},
\
[35]_1 = \overline{A}_{32} \overline{A}_{55}+ \overline{A}_{33} \overline{A}_{54},
\nonumber
\\
&&
C_{4,10,1} =  \left[13^3\right]_1 + \left[3^24\right]_1,
\
\left[13^3\right]_1 =  3A_{11}\overline{A}_{32}^2 \overline{A}_{33}+ A_{12}  \overline{A}_{32}^3/3!,
\nonumber
\\
&&
\left[3^24\right]_1  =  \overline{A}_{32}^2\overline{A}_{44}/2! +
\overline{A}_{33} \overline{A}_{32} \overline{A}_{43},
\nonumber
\\
&&
C_{4,12,1} = \left[3^4\right]_1 = \overline{A}_{32}^3  \overline{A}_{33}/3!.
\nonumber
\end{eqnarray}
\end{small}
By (\ref{456e}), the terms needed for $\Delta_{re}$ for $e=f,g$ and $r=4,5,6$ are
given by (\ref{e12}) for $r=5$, and
\begin{eqnarray}
&&
r =4:\ e_{21}  =  \overline{A}_{23} H_{1} +[4]_1\ H_3 +
\left[3^2\right]_1\ e\left(3^2\right) + [13]_1\ e(13),
\nonumber
\\
&&
r =5:\  e_{31} = \sum\left\{ [\pi]_1\ e(\pi): \pi \in S_3(e)\right\},
\nonumber
\\
&&
r =6:\  e_{22} = \sum\left\{ [\pi]_2\ e(\pi): \pi \in S_2(e)\right\},
\nonumber
\\
&&
e_{41} = \sum\left\{ [\pi]_1\ e(\pi): \pi \in S_4(e)\right\},
\nonumber
\end{eqnarray}
where $S_r(e)$ is given in Appendix C for $e=f,g$.
\end{theorem}

Using Theorem 5.1, $g_{12}$ is given by (\ref{e12}), and
\begin{eqnarray*}
\Delta_{5g}=g_{12}+g_{31},
\end{eqnarray*}
where
\begin{eqnarray*}
g_{31}=\sum\left\{[\pi]_1g(\pi),\ \pi\in S_3(g)\right\},
\
S_3(g)=\left\{5,34,3^3,23\right\}.
\end{eqnarray*}
By Table 1.1, $g_4$ has 8+3 terms, or 2+2 terms for $X_{12}$ and 2+1 terms for $X_{23}$.
For $r\leq 3$,  $\Delta_{re}$ does not depend on $e$.
If $l_{03}=0$, then
\begin{eqnarray*}
\Delta_{2h}-l_{01}A_{12} =\Delta_{2f} = \Delta_{2g} = A_{23}H_1/2 +A_{45}H_3/4!.
\end{eqnarray*}
For $e=f,g$, $e_{22}$ needs $[4]_2=\overline{A}_{45}$, $[2]_2=\overline{A}_{24}$, $[3^2]_2$ above and
\begin{eqnarray*}
[13]_2= A_{11}\overline{A}_{34}+A_{12}\overline{A}_{33}+A_{13}\overline{A}_{32}.
\end{eqnarray*}
For $g_{41}$ needs $[\pi]_1$ for $\pi\in S_4(g)=\{6,4^2,24,2^2,35,3^24,3^4,23^2\}$:
$f_{41}$ needs $[\pi]_1$ for $\pi\in S_4(f)= S_4(g)\cup  \{15,1^24,1^22,134,13^3,123,1^23^2,1^33\}$.

\section{Cumulant coefficients for $Y_{JK\theta}$ of (\ref{JKth})}
\setcounter{equation}{0}

Theorem 6.1 gives $A_{ri\theta}^{JK}$ of (\ref{rith}) in terms of $A_{ri\theta}$ of (\ref{Arith}).
Its proof is outlined as follows:
Write $s=s_{2K\theta}$ of (\ref{rJ}) as $a_{21\theta}n^{-1}(1+\epsilon)$,
where $\epsilon=\sum_{j=1}^{K-1} x_j n^{-j}$, $x_j=a_{2,j+1,\theta}$.
So,
\begin{eqnarray*}
(1+\epsilon)^{-r/2}=\sum_{k=0}^\infty {-r/2\choose k}\epsilon^k.
\end{eqnarray*}
By (\ref{A.1}),
\begin{eqnarray*}
\epsilon^k=\sum_{j=k}^\infty \widehat{B}_{jk}(x) n^{-j}.
\end{eqnarray*}
So,
\begin{eqnarray*}
s^{-r/2}= \left(a_{21\theta} n^{-1}\right)^{-r/2} \sum_{j=0}^\infty d_{rj} n^{-j},
\end{eqnarray*}
where
\begin{eqnarray*}
d_{rj}=\sum_{k=0}^j  {-r/2\choose k} \widehat{B}_{jk}(x)
\end{eqnarray*}
depends on
\begin{eqnarray*}
&&
d_{r0} =1, d_{r1}=(-r/2) x_1,
\
d_{r2}=(-r/2) x_2+{-r/2\choose 2} x_1^2,
\\
&&
x_1 =A_{22\theta}\ I(1<K),
\
x_2=A_{23\theta}\ I(2<K),
\
\cdots .
\end{eqnarray*}
Theorem 6.1 and Corollary 6.1 are now immediate.

\begin{theorem}
With notation as above,
\begin{eqnarray}
&&
A_{1i\theta}^{JK} =0\mbox{ for }i\leq J,
\\
&&
A_{1i\theta}^{JK} = A_{1i\theta}\otimes  d_{1i}=\sum_{j=J+1}^i  d_{1,i-j} A_{1j\theta},
\
i\geq J+1,
\nonumber
\\
&&
A_{ri\theta}^{JK} = A_{ri\theta}\otimes  d_{ri}=\sum_{j=r-1}^i  d_{r,i-j} A_{rj\theta},
\
r\geq \max(2,i-1),
\nonumber
\end{eqnarray}
where $a_i\otimes b_i=\sum_{j=0}^i a_jb_{i-j}$.
\end{theorem}

\begin{corollary}
With notation as above,
\begin{eqnarray*}
A_{1,J+1,\theta}^{JK} = A_{1,J+1,\theta},
\
A_{1,J+2,\theta}^{JK} = A_{1,J+1,\theta}d_{11}+ A_{1,J+2,\theta}
\end{eqnarray*}
and for $r\geq 2$,
\begin{eqnarray}
&&
A_{r,r-1,\theta}^{JK} = A_{r,r-1,\theta},
\
A_{rr\theta}^{JK} =  A_{rr\theta} +d_{r1} A_{r,r-1,\theta},
\nonumber
\\
&&
A_{r,r+1,\theta}^{JK} =  A_{r,r+1,\theta} +  d_{r1} A_{rr\theta}  +d_{r2}A_{r,r-1,\theta}.
\nonumber
\end{eqnarray}
For $r=2$ it is simpler to use
\begin{eqnarray}
A_{2i\theta}^{JK}
&=&
0\mbox{ for }2\leq i\leq K,
\nonumber
\\
&=&
\sum_{j=K+1}^i d_{2,i-j} A_{2j\theta}
\mbox{ for }i>K.
\nonumber
\end{eqnarray}
\end{corollary}

The last result of the corollary follows since $\kappa_2(Y_{JK\theta})-1=
s_{2\infty\theta}/s_{2K\theta}-1$ $=$ $s_{2K\theta}^{-1}$ $\sum_{j=K+1}^\infty$ $a_{2j\theta}n^{-j}$.

\section{Expansions for the density of $Y_{01\theta}$ and its derivatives}
\setcounter{equation}{0}

Withers and Nadarajah (2012)
gave expansions for the density and other derivatives of the distribution.
Theorem 7.1 gives expansions under the extended Cornish and Fisher assumption, $l_r=O(1)$ for $r\geq 1$,
and under the more general assumption (\ref{riY}).

\begin{theorem}
Under the extended Cornish and Fisher assumption, $l_r=O(1)$ for $r\geq 1$,
expansions for the density of $Y_{01\theta}$, $p_n$,
and its derivatives follow from (\ref{h}) and (\ref{hL2}):
\begin{eqnarray}
(-D)^i\ p_{n}(x)  = p(x)\sum_{r=1}^\infty n^{-r/2}\ h_{ir}(x,L),
\
i\geq 0,
\nonumber
\end{eqnarray}
where
\begin{eqnarray}
h_{ir}(x,L) = \sum_{k=1}^{3r} C_{rk} H_{k+i}(x),
\nonumber
\end{eqnarray}
where  $h_{ri}$ has a different meaning from that in (\ref{erxL})-(\ref{3.2}).
Under the more general assumption (\ref{riY}), the expansions take the form
\begin{eqnarray}
(-D)^i\ p_{n}(x)  = p(x)\sum_{r=1}^\infty n^{-r/2}\ h_{ir}(x),
\
i\geq 0,
\nonumber
\end{eqnarray}
where $h_{ir}(x)$ is $h_r(x)$ with $\{H_k\}$ replaced by $\{H_{k+i+1}\}$.
For example,
\begin{eqnarray*}
h_{ir}(x) = {h}_{ir}(x,l)+\Delta_{ri{h}},
\end{eqnarray*}
where
\begin{eqnarray*}
&&
\Delta_{ri{h}}=0\ \mbox{for}\ r=1,2,
\\
&&
\Delta_{3i{h}}=A_{12}H_{i+1}+\overline{A}_{33}H_{i+3},
\\
&&
\Delta_{4i{h}}=\left(A_{11}A_{12}+\overline{A}_{23}\right)
H_{i+2}+(A_{12}\overline{A}_{32}+A_{11}
\overline{A}_{33}+\overline{A}_{44})H_{i+4} +\overline{A}_{32}\overline{A}_{33}H_{i+6}.
\end{eqnarray*}
\end{theorem}

This follows from its Charlier expansion version given in equation (2.6)
of Withers and Nadarajah (2012), so it remains valid for general $X$.
See equation (4.6) there for the case $X=N$.
For its multivariate extension, see Section 7 of Withers and Nadarajah (2011).

\section*{Appendix A: The ordinary Bell polynomials}
\renewcommand{\theequation}{$\mbox{A.\arabic{equation}}$}
\setcounter{equation}{0}

The {\it ordinary Bell polynomial}, $\widehat{B}_{rj}(y)$,
is defined in terms of a sequence $y=(y_1,y_2, \ldots)$, by
\begin{eqnarray}
S(t)^j =  \sum_{r=j}^\infty \widehat{B}_{rj}(y)\ t^r
\label{A.1}
\end{eqnarray}
for $j=0,1, \ldots$, where
\begin{eqnarray}
S(t) =\sum_{r=1}^\infty y_r t^r.
\nonumber
\end{eqnarray}
In fact, for $r>0$, $\widehat{B}_{rj}(y)$ is only a function of $\{x_i,\ 1\leq i\leq r-j+1\}$.
For example,
\begin{eqnarray*}
&&
\widehat{B}_{rj}(y) =0 \mbox{ for }r<j,
\
\widehat{B}_{r0}(y) = I(r = 0),
\\
&&
\widehat{B}_{rj}(-y) = (-1)^j \widehat{B}_{rj}(y),
\\
&&
\widehat{B}_{r1}(y) =y_r,
\
\widehat{B}_{rr}(y)=y_1^r,
\
\widehat{B}_{r+1,r}(y) =r y_1^{r-1}y_2,
\\
&&
\widehat{B}_{r+2,r}(y) =r y_1^{r-1}y_3 +{r\choose 2} y_1^{r-2}y_2^2.
\end{eqnarray*}
They are tabled on page 309 of Comtet (1974) for  $1\leq j\leq r\leq 10$.
Setting
\begin{eqnarray*}
\left[1^{i_1}2^{i_2}\cdots\right]=\left(y_1^{i_1}/i_1!\right)\ \left(y_2^{i_2}/i_2!\right)\ \cdots,
\end{eqnarray*}
we can write $\widehat{B}_{rj}$ and ${b}_{rj}$  of (\ref{rr+2i}) as
\begin{eqnarray}
{b}_{rj}(y)=\widehat{B}_{rj}(y)/j!=\sum \left[1^{i_1}2^{i_2}\cdots\right]
\label{B}
\end{eqnarray}
summed over all partitions of $r$, that is, over $1{i_1}+2{i_2}+\cdots=r$, subject to $i_1+i_2+\cdots=j$.

Taking the coefficient of $t^r$ in $S(t)^{j+k}=S(t)^{j}S(t)^{k}$ gives
the recurrence formula
\begin{eqnarray*}
\widehat{B}_{r,j+k}(y) =\widehat{B}_{rj}(y)\otimes \widehat{B}_{rk}(y)
=\sum_{a+b=r} \widehat{B}_{aj}(y) \widehat{B}_{bk}(y).
\end{eqnarray*}
For example, $k=1$ gives
\begin{eqnarray*}
\widehat{B}_{r,j+1}(y) = \sum_{a=j}^{r-1} \widehat{B}_{aj}(y) y_{r-a},
\
r\geq j+1,
\end{eqnarray*}
and $j=k=1$ gives
\begin{eqnarray*}
{b}_{r2}(y)=\widehat{B}_{r2}(y/2)
&=&
\sum_{a=1}^{r-1} y_a y_{r-a}/2,
\
r\geq 2,
\\
&=&
\begin{cases}
\displaystyle
\sum_{a=1}^{b} y_a y_{r-a}, & \mbox{ if } r = 2b+1,
\\
\displaystyle
x_b^2/2 +  \sum_{a=1}^{b-1} y_a y_{r-a}, & \mbox{ if } r= 2b.
\end{cases}
\end{eqnarray*}
The {\it exponential Bell polynomial}, ${B}_{rj}(x)$,
is the coefficient of $t^r/r!$ in $S(t)^j/j!$ when $y_j=x_j/j!$.
So,
\begin{eqnarray*}
{B}_{rj}(x)/r!=\widehat{B}_{rj}(y)/j! = {b}_{rj}(y).
\end{eqnarray*}
They are tabled on pages 307, 308 of Comtet (1974) for  $1\leq j\leq r\leq 12$.

It is easy to show that if $Y_r= a b^r y_r$, $X_r= a b^r x_r$ then
\begin{eqnarray}
\widehat{B}_{rj}(Y)= a^jb^r \widehat{B}_{rj}(y),
\
{B}_{rj}(X)= a^jb^r {B}_{rj}(x).
\label{trans}
\end{eqnarray}
The multinomial expansion can be written as $S(1)^r/r!=\sum [1^{i_1}2^{i_2}\cdots]$ summed over all partitions of $r$.

\section*{Appendix B: Derivatives of $H_{r}$, and $H_r$ as a function of $\bf{ a}$}
\renewcommand{\theequation}{$\mbox{B.\arabic{equation}}$}
\setcounter{equation}{0}

We first show that $H_r$ has $k$th derivative
\begin{eqnarray}
H_{r \cdot k} =\sum_{i=0}^k {k \choose i} (-1)^i b_{k-i}H_{r+i},
\
k\geq 0,
\label{r.k}
\end{eqnarray}
where
\begin{eqnarray}
b_{i} = \left(H_1+D\right)^i 1 = \left(H_1+D\right)b_{i-1}.
\nonumber
\end{eqnarray}
In particular,
\begin{eqnarray}
&&
b_0 =1,
\
b_1=H_1,
\
b_2 =2H_1^2-H_2,
\
b_{3}= 3!H_1^3-6H_1H_2+H_3,
\nonumber
\\
&&
b_{4} = 4!H_1^4-36 H_1^2H_2 +8H_1H_3+6H_2^2-H_4,
\nonumber
\\
&&
b_{5} = 5!H_1^5-240 H_1^3H_2 +60H_1^2H_3+90H_1H_2^2-10H_1H_4-20H_2H_3 +H_5.
\nonumber
\end{eqnarray}
We also prove (\ref{arH}), giving $b_{k}$ in terms of ${\bf a}$.

Rewriting (\ref{recur}) as
\begin{eqnarray}
H_{r \cdot 1} =  H_1 H_r-H_{r+1},
\
r\geq 0,
\nonumber
\end{eqnarray}
we obtain
\begin{eqnarray}
H_{r \cdot k} =\sum_{i=0}^k \alpha_{ki}H_{r+i},
\
k\geq 0,
\nonumber
\end{eqnarray}
where
\begin{eqnarray}
\alpha_{k+1,i} = \left(H_1+D\right) \alpha_{ki} - \alpha_{k,i-1},
\nonumber
\end{eqnarray}
where $\alpha_{ki}=0$ if $i<0$.
The second equation follows from $H_{r \cdot k+1}=DH_{r \cdot k}$.
This gives
\begin{eqnarray*}
&&
\alpha_{kk} = (-1)^k,
\
\alpha_{k,k-1}=(-1)^{k-1} kH_1,
\\
&&
\alpha_{k,k-i} = (-1)^{k-i} {k\choose i}b_{i},
\
b_{i} = \alpha_{i0}.
\nonumber
\end{eqnarray*}
But by (\ref{recur}),
\begin{eqnarray}
H_r
&=&
\left(a_1-D\right)H_{r-1} = \left(a_1-D\right)^r\ 1,
\
r \geq 1,
\nonumber
\\
&=&
(-1)^r B_r(-{\bf a}) =\sum_{i=0}^r (-1)^{r-i} B_{ri}({\bf a}),
\
r\geq 0.
\label{Hra}
\end{eqnarray}
In particular,
\begin{eqnarray}
&&
H_1 = a_1,
\
H_2= a_1^2 -a_2,
\
H_3 =  a_1^3-3a_1a_2+a_3,
\nonumber
\\
&&
H_4 = a_1^4 -6a_1^2a_2+3a_2^2 +4a_1a_3 -a_4,
\label{4a}
\\
&&
H_5 =a_1^5 -10a_1^3a_2 +15a_1 a_2^2+10a_1^2a_3-10a_2a_3 -5a_1a_4+a_5,
\label{5a}
\\
&&
H_6 =a_1^6 -15a_1^4a_2 +45a_1^2 a_2^2 -15a_2^3
+20a_1^3a_3 -60a_1a_2a_3 +10a_3^2
\nonumber
\\
&&
-15a_1^2a_4 +15a_2a_4+6a_1a_5-a_6-10a_2a_3+10a_1^2a_3  -10a_1^3a_2+a_1^5.
\label{6a}
\end{eqnarray}
So, Comtet (1974)'s table gives $H_r$ in terms of ${\bf a}$ to $r=12$.
Replacing ${\bf a}$ by  $-{\bf a}$ gives
\begin{eqnarray*}
b_{r}= B_r({\bf a}) =\sum_{i=0}^r  B_{ri}({\bf a}),
\
r\geq 0,
\end{eqnarray*}
proving (\ref{arH}).
$B_r({\bf a})$ is called {\it the $r$th complete exponential Bell polynomial}.
So, Comtet (1974)'s table pages 307-308 gives $b_{r}$ in terms of $ {\bf a}$ up to $r=12$.
The first six are
\begin{eqnarray*}
&&
b_{0} = 1,
\
b_{10}= a_1,
\
b_{2}= a_2+a_1^2,
\
b_{3}=a_3+3a_1a_2+a_1^3,
\\
&&
b_{4} = a_4 +4a_1a_3+3a_2^2 +6a_1^2a_2 +a_1^4,
\\
&&
b_{5} = a_5 +5a_1a_4+10a_2a_3 +10a_1^2a_3+15a_1a_2^2 +10a_1^3a_2 +a_1^5,
\\
&&
b_{6} = a_6 +6a_1a_5+15a_2a_4+10a_3^2 +15a_1^2a_4+60a_1a_2a_3+15a_2^3
\nonumber
\\
&&
+20a_1^3a_3 +45a_1^2a_2^2 +15a_1^4a_2 +a_1^6.
\end{eqnarray*}
For example, this gives
the first four derivatives of $H_r$ in terms of $\alpha_{44}$, $\alpha_{43}$ above and
$\alpha_{42} =6b_{2}$, $\alpha_{41}=-4b_{3}$.

\noindent
{\bf Expressions for $a_r$ in terms of ${\bf H}$:} By (\ref{r.k}),
\begin{eqnarray}
a_r = H_{1 \cdot r-1} =\sum_{i=0}^{r-1}{r-1\choose i} (-1)^i b_{r-1-i}H_{1+i},
\
r\geq 1.
\nonumber
\end{eqnarray}
In particular,
\begin{eqnarray}
&&
a_1 = H_1,
\
a_2 =H_1^2-H_2,
\
a_3= 2H_1^3-3H_1H_2+H_3,
\nonumber
\\
&&
a_4  = 3!H_1^4-12 H_1^2H_2 +4H_1H_3+3H_2^2-H_4,
\nonumber
\\
&&
a_5 = 4!H_1^5 -60H_1^3H_2 +20 H_1^2H_3 +30H_1H_2^2-5H_1H_4-10H_2H_3+H_5,
\nonumber
\\
&&
a_6 = 5!H_1^6 -360H_1^4H_2 +120 H_1^3H_3 -30H_1^2H_4+6 H_1H_5 -H_6 +270H_1^2H_2^2
\nonumber
\\
&&
-120H_1H_2H_3 -30H_2^3 +15H_2H_4.
\nonumber
\end{eqnarray}
A simpler way to express  $a_r$ in terms of $\{H_i\}$ is to apply
Faa di Bruno's rule, [4i] of Comtet (1974), to obtain
the $r$th derivative of $a(x)= f(p(x))$ at $f(p)=-\ln p$, in
terms of ${\bf p}=(p_1,p_2, \ldots)$, where $p_r= D^r p(x)$:
\begin{eqnarray*}
a_{r}=\sum_{j=1}^r f_j B_{rj}({\bf p}),
\
r\geq 1,
\end{eqnarray*}
where $f_j= F_j (-p)^{-j}$ at $F_j=(j-1)!$, $p=p(x)$ is the $j$th
derivative of $f(p)$ at $p$.
But  $p_r= a b^r H_r$, where $b=-1$, $a=p$.
So, by (\ref{trans}),  $B_{rj}({\bf p})=p^j (-1)^r B_{rj}({\bf H})$.
This gives the simple inverse formula
\begin{eqnarray}
a_r =  \sum_{j=1}^r (-1)^{r-j} (j-1)! B_{rj}({\bf H}),
r\geq 1.
\nonumber
\end{eqnarray}
So, Comtet (1974)'s table pages 307-308 gives $a_r$ in terms of ${\bf H}$ up to $r=12$.
For example,
\begin{eqnarray*}
a_6 =-B_{61}+B_{62}-2B_{63}+3!B_{64}-4!B_{65}+5!B_{66},
\end{eqnarray*}
where
\begin{eqnarray*}
&&
B_{61} =H_6, B_{62}=6H_1H_5+15H_2H_4,
\
B_{63}=15H_1^2H_4+60H_1H_2H_3+15H_2^3,
\\
&&
B_{64} = 20H_1^3H_3+45H_1^2H_2^2,
\
B_{65}=15H_1^4H_2,
\
B_{66}=H_1^6.
\end{eqnarray*}
This is the inverse formula to (\ref{Hra}).
These two relations are
essentially the relations between the non-central moments and the cumulants:
$-a=\ln p$, $p=e^{-a}$, $K=\ln M$, $M=e^K$,
where $M$, $K$ are the moment and cumulant generating functions.
So, $-a_r$ can be identified with the $r$th cumulant, and $H_r$ with  the $r$th moment.

An alternative way to express $b_{r}$ in terms of $\{a_i\}$ of (\ref{ar}), is to set
\begin{eqnarray*}
e_1 =0,
\
e_r=\sum_{i=2}^r {r\choose i}a_1^{r-i}a_{i}\mbox{ if }r\geq 2,
\
\delta_r  = b_{r}-a_1^r-e_r.
\end{eqnarray*}
Then,
\begin{eqnarray*}
\delta_{2} = \delta_{3}=0,
\
\delta_{r+1} = \left(a_1+D\right)\delta_{r} +ra_{2}e_{r-1},
\
r\geq 1.
\end{eqnarray*}
This gives
\begin{eqnarray*}
&&
b_{r} =a_1^r+e_r+\delta_r,
\
r\leq 6,
\\
&&
\delta_{4}  =3a_{2}^2,
\\
&&
\delta_5 = 15 a_1a_{2}^2+10 a_{2}a_{3},
\\
&&
\delta_6 = 45 a_1^2a_{2}^2+60 a_1a_{2}a_{3}+15a_{2}^3+15a_{2}a_{4} +10a_{3}^2.
\end{eqnarray*}

\section*{Appendix C: $e_r(x,L)$ for $e=f,g$ in terms of ${\bf H}$}
\renewcommand{\theequation}{$\mbox{C.\arabic{equation}}$}
\setcounter{equation}{0}

(\ref{fgpi}) gives $f_r(x,L)$, $g_r(x,L)$, $r\leq 6$ in terms of
certain $e(\pi)$.
We now give these in terms of ${\bf H}$.
First consider the case  $l_3=0$.
Let $S_{0r}(e)$  denote
the partitions in $S_{r}(e)$ needed when $l_3= 0$.
For each $r$ we first give $e(\pi)$ covered by the special cases (\ref{hfg}), (\ref{1k+1}), (\ref{kapm}).
\begin{small}
\begin{eqnarray*}
&&
r =2:\ S_{02}(f) = \{4,2\},
\
f(4) =H_{3},
\
f(2)=H_{1}.
\\
&&
r =3:\
S_{03}(f) = \{5,14,12\},
\
f(5) =H_{4},
\
f(12) = H_2-H_1^2=-H_{1 \cdot 1} \mbox{ of }(\ref{1k+1}), (\ref{kapm}),
\\
&&
f(14) = H_4-H_1H_3 = -H_{3 \cdot 1}.
\\
&&
r =4:\
S_{04}(f) = \left\{ 6,4^2,24,2^2,15,1^2 4, 1^2 2 \right\},
\
f(6) =H_5,
\
f\left(1^22\right) \mbox{ of }(\ref{kapm}),
\\
&&
f\left(4^2\right)  = H_7-H_1 H_3^2,
\\
&&
f(24)  = H_5-H_1^2H_3,
\
f\left(2^2\right) = H_3- H_1^3,
\
f(15) =H_5-H_1H_4=-H_{4 \cdot 1} \mbox{ of }(\ref{1k+1}),
\\
&&
f\left(1^2 4\right) =H_5-H_2H_3 -2H_1H_4 +2H_1^2H_3.
\\
&&
r =5:\
S_{05}(f) = \left\{7,45,25, 16,14^2,124,12^2,1^25,1^34,1^32\right\},
\
f(7) =H_6,
\
f\left(1^32\right) \mbox{ of }(\ref{kapm}),
\\
&&
f(45) = H_8-H_1H_3H_4,
\
f(25) = H_6-H_1^2H_4,
\
f(16) = H_6 -H_1H_5=-H_{5 \cdot 1} \mbox{ of }(\ref{1k+1}),
\\
&&
f\left(14^2\right) =  H_8-H_1H_7 -2H_1H_3H_4 -H_2H_3^2+3H_1^2H_3^2,
\\
&&
f(124) =H_6 -H_1H_5 -2H_1H_2H_3-H_1^2H_4 +3H_1^3H_3,
\\
&&
f\left(12^2\right) = H_4 -H_1H_3 -3H_1^2H_2 +3H_1^4,
\\
&&
f\left(1^2 5\right) = H_6-H_2H_4-2H_1H_5+2H_1^2H_4,
\\
&&
f\left(1^3 4\right) = H_6 -3H_1H_5 -3H_2H_4 +6H_1^2H_4-H_3^2+6H_1H_2H_3 -6H_1^3H_3.
\end{eqnarray*}
\end{small}
For $r=6$, because of the increasing number of terms, we adapt the notation
of Comtet (1974), setting
\begin{eqnarray}
k \cdot 1^{i_1}2^{i_2}\cdots \stackrel{H}{=} kH_1^{i_1}H_2^{i_2}\cdots
\label{com}
\end{eqnarray}
when giving formulas for $f(\pi)$, $g(\pi)$.
For example,
\begin{eqnarray*}
5 \cdot 1^46 &\stackrel{H}{=} 5H_1^4H_6,
\
15 \cdot 1^32^2\stackrel{H}{=}15H_1^3H_2^2,
\
3 \cdot 4(11)\stackrel{H}{=}3H_4H_{11}.
\end{eqnarray*}
Using this notation, when $l_3=0$, $f_6(x,L)$ is given by
\begin{small}
\begin{eqnarray*}
&&
r =6:
\\
&&
S_{06}(f) =  \left\{8, 5^2,46,4^3,26,24^2, 2^24, 2^3, 17,145,125,1^26,1^2 4^2,1^22^2,1^35,1^44,1^42,1^6\right\},
\\
&&
f(8) =H_7 \stackrel{H}{=}{7},
\
f\left(1^42\right) \mbox{ of }(\ref{kapm}),
\
f\left(5^2\right) \stackrel{H}{=} 9-14^2,
\
f(46) \stackrel{H}{=}9-135,
\\
&&
f\left(4^3\right) \stackrel{H}{=}11-3 \cdot 137-23^3+3 \cdot 1^23^3,
\\
&&
f(26) \stackrel{H}{=}7-1^25,
\\
&&
f\left(24^2\right) \stackrel{H}{=} 9-1^27-2 \cdot 135 -123^2+3 \cdot 1^33^2,
\\
&&
f\left(2^2 4\right) \stackrel{H}{=}7-2 \cdot 1^25  -13^2 -1^223 +3 \cdot 1^43,
\\
&&
f\left(2^3\right) \stackrel{H}{=}5 -3 \cdot 1^23-1^32 +3 \cdot 1^5,
\\
&&
f(17) \stackrel{H}{=} 7 -16=-H_{6 \cdot 1} \mbox{ of }(\ref{1k+1}),
\\
&&
f(145) \stackrel{H}{=} 9 -18 -135-14^2 -234 +3 \cdot 1^234,
\\
&&
f(125) \stackrel{H}{=} 7 -16 -1^25-2 \cdot 124 +3 \cdot 1^34,
\\
&&
f\left(1^26\right) \stackrel{H}{=} 7 - 2 \cdot 16 -25 +2 \cdot 1^25,
\\
&&
f\left(1^2 4^2\right) \stackrel{H}{=} 9 -2 \cdot 18 -27 +2 \cdot 1^27  -
2 \cdot 135 -2 \cdot 14^2-4 \cdot 234 +12 \cdot 1^234
\\
&&
-3^3 +9 \cdot 123^2 -12 \cdot 1^33^2,
\\
&&
f\left(1^2 24\right) \stackrel{H}{=}7-2 \cdot 16 -25 +1^25-4 \cdot 124+
6 \cdot 1^34 -2 \cdot 13^2 -2 \cdot 2^23 +15 \cdot 1^223 -12 \cdot 1^43,
\\
&&
f\left(1^2 2^2\right) \stackrel{H}{=} 5-2 \cdot 14 -23 -1^23 -6 \cdot 12^2 +21 \cdot 1^32  -12 \cdot 1^5,
\\
&&
f\left(1^3 5\right) \stackrel{H}{=} 7 -3 \cdot 16-3 \cdot 25 +6 \cdot 1^25-34 +6 \cdot 124 -6 \cdot 1^34,
\\
&&
f\left(1^4 4\right) \stackrel{H}{=}  7  - 4 \cdot 16 -6 \cdot 25 +12 \cdot 1^25 -5 \cdot 34 +24 \cdot 124 -24 \cdot 1^34 +8 \cdot 13^2  +6 \cdot 2^23
\\
&&
-36 \cdot 1^223 +24 \cdot 1^43.
\end{eqnarray*}
\end{small}
A second special case for $g(\pi)$ is
\begin{eqnarray}
g(2k) = H_{k+1}-H_1H_k-H_{k-1}\left(H_2-H_1^2\right),
\
k\geq 1.
\label{g2k}
\end{eqnarray}
For example, $g(24) =H_5-H_1H_4-H_2H_3 +H_1^2H_3$.
When $l_3=0$, $g_r(x,L)$ is given by (\ref{fgpi}) as follows:
\begin{small}
\begin{eqnarray*}
&&
r =2:
\
S_{02}(g) = \{4,2\},
\
g(4)=H_{3},
\
g(2)=H_{1}.
\\
&&
r =3:
\
S_{03}(g) = \{5\},
\
g(5)=H_{4}.
\\
&&
r =4:
\
S_{04}(g) = \left\{6,4^2,24,  2^2\right\},
\
g(6) =H_5,
\
g(24), g\left(2^2\right)  \mbox{ of  (\ref{g2k})},
\\
&&
g\left(4^2\right) =H_7-2H_3H_4 +H_1 H_3^2.
\\
&&
r =5:\
S_{05}(g) = \{25,45,7\},
\
g(7)=H_6,
\
g(25) \mbox{ of (\ref{g2k})},
\\
&&
g(45) = H_8-H_3H_5-H_4^2+H_1H_3H_4.
\\
&&
r =6:
\
S_{06}(g) = \left\{8,5^2,46,4^3,26,2 4^2,2^2 4,2^3\right\},
\
g(8) =H_7 \stackrel{H}{=}{7},
\
g(26) \mbox{ of (\ref{g2k})},
\\
&&
g\left(5^2\right) \stackrel{H}{=} 9 -2 \cdot 45 +14^2,
\\
&&
g(46) \stackrel{H}{=}9-36 -45+135,
\\
&&
g\left(4^3\right) \stackrel{H}{=}11-3 \cdot 38-3 \cdot 47+3 \cdot 137+3 \cdot 3^25+6 \cdot 34^2-9 \cdot 13^24-23^3+3 \cdot 1^23^3,
\\
&&
g\left(24^2\right) \stackrel{H}{=} 9-18-27 +1^27-2 \cdot 36-2 \cdot 45 +4 \cdot 135
+2 \cdot 14^2+4 \cdot 234-6 \cdot 1^234
\\
&&
+3^3-4 \cdot 123^2+3 \cdot 1^33^2,
\\
&&
g\left(2^2 4\right) \stackrel{H}{=}7-2 \cdot 16 -2 \cdot 25 +3 \cdot 1^25 -2 \cdot 34 +4 \cdot 124 -3 \cdot 1^34
\\
&&
+3 \cdot 13^2+2 \cdot 2^23-7 \cdot 1^223 +3 \cdot 1^43,
\\
&&
g\left(2^3\right) \stackrel{H}{=}5 -3 \cdot 14 -3 \cdot 23 +6 \cdot 1^23 +6 \cdot 12^2-10 \cdot 1^32 +3 \cdot 1^5.
\end{eqnarray*}
\end{small}
When $l_3=0$,
\begin{eqnarray}
h_2(x,L) - \left[1^2\right]H_1=f_2(x,L)=g_2(x,L) = [2]H_1+[4]H_3.
\nonumber
\end{eqnarray}
For the case $X=N$, it is known that  $g_r(x,L) - I (r = 1) L_1 $
does not depend on $L_1$.
This is a key step  used in  of
Withers (1989a, 1983) to construct parametric and non-parametric
confidence intervals of level say $0.95 +O(n^{-r/2})$, given $r\geq 1$.
We have seen that this property is also true for general $X$ for $r\leq 6$.
We now show that it is true for all $r$.
Set
\begin{eqnarray*}
s=\overline{P}_n(y)=\Pr\left(Y_{01\theta}-\lambda_{1n}\leq y\right)=P_n\left(y+\lambda_{1n}\right)=P(x)
\end{eqnarray*}
say.
Setting $g_0(x)=x$ gives
\begin{eqnarray*}
\sum_{r=0}^\infty n^{-r/2} g_r(x)=
P_n^{-1}(s)=\lambda_{1n}+y=\lambda_{1n}+\overline{P}_n^{-1}(s)
= n^{-1/2}l_1+ \sum_{r=0}^\infty n^{-r/2} \left[g_r(x)\right]_{l_1=0}.
\end{eqnarray*}
Taking the coefficient of $n^{-r/2}$ gives $g_r(x,L) = I (r = 1) L_1 +[g_r(x)]_{l_1=0}$.

We now give the extra terms needed when  $l_3\neq 0$.
\begin{small}
\begin{eqnarray*}
&&
r =2:
\\
&&
f\left(3^2\right) = H_5-H_1 H_2^2,
\
f(13) =H_3-H_1H_2=- H_{2 \cdot 1} \mbox{ of }(\ref{1k+1}).
\\
&&
r = 3:
\\
&&
f(34) = H_6 - H_1 H_2H_3,
\\
&&
f\left(3^3\right)  =H_8 - 3H_1 H_2H_5 -H_2^4+3 H_1^2H_2^3,
\\
&&
f(23)  = H_4 - H_1^2H_2,
\\
&&
f\left(13^2\right)  =H_6 -H_1H_5-2H_1H_2H_3-H_2^3 + 3H_1^2H_2^2,
\\
&&
f\left(1^2 3\right)  =H_4 -2 H_1H_3-H_2^2 +2H_1^2H_2.
\end{eqnarray*}
\end{small}
\begin{small}
\begin{eqnarray*}
&&
r=4:
\\
&&
f(35) \stackrel{H}{=} 7-1 2 4,
\\
&&
f\left(3^2 4\right)  =  9-2 \cdot 126 -135 -2^33 +3 \cdot 1^22^23,
\\
&&
f\left(3^4\right)  \stackrel{H}{=} 11  -4 \cdot 128 -3 \cdot 15^2 -6 \cdot 2^35 +18 \cdot 1^2 2^25 -2^43 +10 \cdot 12^ 5 -15 \cdot  1^3 2^4,
\\
&&
f\left(23^2\right) \stackrel{H}{=}7 -2 \cdot 124 -12^3  -1^25 +3 \cdot 1^32^2,
\\
&&
f(134)  \stackrel{H}{=}7 -16 -124 -13^2-2^23 +3 \cdot 1^223,
\\
&&
f\left(1 3^3\right) \stackrel{H}{=}9  -18 -3 \cdot 126 -3 \cdot 135-3 \cdot 2^25 +9 \cdot 1^225 -4 \cdot 2^33
\\
&&
+9 \cdot 1^22^23  +10 \cdot 12^4  -15 \cdot 1^32^3,
\\
&&
f(123) \stackrel{H}{=} 5-14-1^23 -2 \cdot 12^2 +3 \cdot 1^32,
\\
&&
f\left(1^2 3^2\right) \stackrel{H}{=}7 -2 \cdot 16  -25 +2 \cdot 1^25  -2 \cdot 124 -2 \cdot 13^2 -5 \cdot 2^23
\\
&&
+9 \cdot 12^3 +12 \cdot 1^223 -12 \cdot 1^32^2,
\\
&&
f\left(1^33\right) \stackrel{H}{=}5 -3 \cdot 14-4 \cdot 23  +6 \cdot 1^23+6 \cdot 12^2-6 \cdot 1^32.
\end{eqnarray*}
\end{small}
\begin{small}
\begin{eqnarray*}
&&
r=5:
\\
&&
f(36) \stackrel{H}{=}8-125,
\\
&&
f\left(34^2\right) \stackrel{H}{=}10-2 \cdot 136-127-2^23^2+3 \cdot 1^223^2,
\\
&&
f\left(3^25\right) \stackrel{H}{=}10-2 \cdot 127-145+3 \cdot 1^22^24-2^34,
\\
&&
f\left(3^34\right) \stackrel{H}{=}12-3 \cdot 129-138-3 \cdot 156+9 \cdot 1^22^26-3 \cdot 2^36-2^33^2+9 \cdot 1^2235-3 \cdot 2^235
\\
&&
+10 \cdot 12^43-15 \cdot 1^32^33,
\\
&&
f\left(3^5\right) \stackrel{H}{=} (14)-5 \cdot 12(11) -10 \cdot 158 -10 \cdot 2^38 +30 \cdot 1^22^28
\\
&&
-15 \cdot 2^25^2 +45 \cdot 1^225^2-10 \cdot 2^335
\\
&&
+100 \cdot 12^45 -150 \cdot 1^32^35 -2^54+15 \cdot 12^53 +10 \cdot 2^7-105 \cdot 1^22^6+105 \cdot 1^42^5,
\\
&&
f(234) \stackrel{H}{=} 8-1^26-125-134 -12^23+3 \cdot 1^323,
\\
&&
f\left(23^3\right) \stackrel{H}{=} 10-1^28-3 \cdot 127-3 \cdot 145+9 \cdot 1^325-3 \cdot 12^25+9 \cdot 1^22^24
\\
&&
-3 \cdot 2^34-12^33 +10 \cdot 1^22^4-15 \cdot 1^42^3,
\\
&&
f\left(2^2 3\right) \stackrel{H}{=}6 -123-2 \cdot 1^24-1^22^2 +3 \cdot 1^42,
\\
&&
f(135) \stackrel{H}{=} 8 -17-125 -134 -2^24 +3 \cdot 1^224,
\\
&&
f\left(13^24\right) \stackrel{H}{=} (10) -19 -2 \cdot 127  -3 \cdot 136 -2 \cdot 2^26 +6 \cdot 1^226 -145
\\
&&
+3 \cdot 1^235 -235  -2^34 +3 \cdot 1^22^24
\\
&&
-3 \cdot 2^23^2 +6 \cdot 1^223^2 +10 \cdot 12^33 -15 \cdot 1^32^23,
\\
&&
f\left(13^4\right) \stackrel{H}{=}(12) -1(11)  -4 \cdot 129 -4 \cdot 138-4 \cdot 2^28+12 \cdot 1^228
\\
&&
-6 \cdot 156-6 \cdot 2^36+18 \cdot 1^22^26
\\
&&
-3 \cdot 25^2+9 \cdot 1^25^2-18 \cdot 2^235+36 \cdot 1^2235+60 \cdot 12^35-90 \cdot 1^32^25 -2^44 -4 \cdot 2^33^2
\\
&&
+55 \cdot 12^43-60 \cdot 1^32^33+10 \cdot 2^6-105 \cdot 1^22^5+105 \cdot 1^42^4,
\\
&&
f\left(123^2\right) \stackrel{H}{=} 8 -17 -4 \cdot 125+3 \cdot 1^35-2 \cdot 134-2 \cdot 2^24+6 \cdot 1^224
\\
&&
+13 \cdot 1^22^3-1^26-15 \cdot 1^42^2
\\
&&
+6 \cdot 1^323-3 \cdot 12^23 -2^4,
\\
&&
f\left(1^2 34\right) \stackrel{H}{=}8-2 \cdot 17 -26 +2 \cdot 1^26 -125-2 \cdot 2^24 -3 \cdot 134
\\
&&
+6 \cdot 1^224-3 \cdot 23^2 +6 \cdot 1^23^2
\\
&&
+9 \cdot 12^23-12 \cdot 1^323,
\\
&&
f\left(1^2 3^3\right) \stackrel{H}{=} (10) - 2 \cdot 19   -28 + 2 \cdot 1^28 - 3 \cdot 127 - 6 \cdot 136
\\
&&
-6 \cdot 2^26 + 18 \cdot 1^226- 3 \cdot 145  - 9 \cdot 235
\\
&&
+ 18 \cdot 1^235  + 27 \cdot 12^25 - 36 \cdot 1^325 - 4 \cdot 2^34+ 9 \cdot 1^22^24 - 12 \cdot 2^23^2
\\
&&
+ 18 \cdot 1^223^2  + 74 \cdot 12^33
\\
&&
- 90 \cdot 1^32^23 + 10 \cdot 2^5 - 95 \cdot 1^22^4 + 90 \cdot 1^42^3,
\\
&&
f\left(1^2 2 3\right) \stackrel{H}{=}  6-2 \cdot 15-24+1^24-6 \cdot 123+6 \cdot 1^33-2 \cdot 2^3 +15 \cdot 1^22^2 -12 \cdot 1^42,
\\
&&
f\left(1^3 4\right) \stackrel{H}{=}  6-3 \cdot 15-3 \cdot 24+6 \cdot 1^24-3^2 +6 \cdot 123-6 \cdot 1^33,
\\
&&
f\left(1^3 3^2\right) \stackrel{H}{=} 8   - 3 \cdot 17 - 3 \cdot 26 +6 \cdot 1^26 -35 + 4 \cdot 125 - 6 \cdot 1^35
\\
&&
-6 \cdot 134 - 7 \cdot 2^24 + 18 \cdot 1^224
\\
&&
-12 \cdot 23^2 + 18 \cdot 1^23^2  + 66 \cdot 12^23 - 72 \cdot 1^323+ 9 \cdot 2^4- 72 \cdot 1^22^3 + 60  \cdot 1^42^2,
\\
&&
f\left(1^4 3\right) \stackrel{H}{=}  6  - 4 \cdot 15 - 7 \cdot 24 + 12 \cdot 1^24 - 4 \cdot 3^2 + 32 \cdot 123
\\
&&
-24 \cdot 1^33 + 6 \cdot 2^3 - 36 \cdot 1^22^2  + 24 \cdot 1^42.
\end{eqnarray*}
\end{small}
We skip the extra twenty five terms needed for  $f_6$ when $l_3\neq0$.
The extra terms needed for $g_r(x,L)$ in (\ref{fgpi}) when $l_3\neq0$ are as follows:
\begin{small}
\begin{eqnarray*}
&&
r =2:
\\
&&
g\left(3^2\right) = H_5-2H_2H_3 +H_1 H_2^2.
\\
&&
r =3:
\\
&&
g(34) = H_6 -H_2H_4 -H_3^2 +H_1 H_2H_3,
\\
&&
g\left(3^3\right) = H_8 - 3H_2H_6 -3H_3H_5 +3H_1H_2H_5 +3H_2^2H_4+6H_2H_3^2
\\
&&
-9H_1H_2^2H_3 -H_2^4+3H_1^2H_2^3,
\\
&&
g(23) = H_4-H_1H_3 - H_2^2 + H_1^2H_2.
\\
&&
r =4:
\\
&&
g(35) = 7 -25 - 34 +124,
\\
&&
g\left(3^2 4\right)  \stackrel{H}{=} 9-2 \cdot 27-3 \cdot 36 +2 \cdot 126-45 +135 +2^25+6 \cdot 234
\\
&&
+2 \cdot 3^3 -6 \cdot 123^2 -2^33
\\
&&
-3 \cdot 12^24+3 \cdot 1^22^23,
\\
&&
g\left(3^4\right)  \stackrel{H}{=}11-4 \cdot 29 -4 \cdot 38 +4 \cdot 128+6 \cdot 2^27- 6 \cdot 56+24 \cdot 236 -18 \cdot 12^26 +3 \cdot 15^2
\\
&&
+12 \cdot 245 +12 \cdot 3^25 -36 \cdot 1235 -10 \cdot 2^35+18 \cdot 1^22^25 -36 \cdot 2^234 +24 \cdot 12^34
\\
&&
-24 \cdot 23^3 +72 \cdot 12^23^2 +17 \cdot 2^43  -60 \cdot 1^22^33-10 \cdot 12^5 +15 \cdot 1^32^4,
\\
&&
g\left(23^2\right)    \stackrel{H}{=}   7 -16-3 \cdot 25+1^25-2 \cdot 34 +4 \cdot 124 +2 \cdot 13^2 +5 \cdot 2^23
\\
&&
-6 \cdot 1^223 -4 \cdot 12^3   +3 \cdot 1^32^2.
\end{eqnarray*}
\end{small}
\begin{small}
\begin{eqnarray*}
&&
r =5:
\\
&&
g(36) \stackrel{H}{=}8-26-35+125,
\\
&&
g\left(34^2\right) \stackrel{H}{=}  10-28-3 \cdot 37+127-2 \cdot 46+2 \cdot 136+2 \cdot 235 +2 \cdot 24^2
\\
&&
+5 \cdot 3^24-6 \cdot 1234-3 \cdot 13^3 - 2^23^2 +3 \cdot 1^223^2,
\\
&&
g\left(3^25\right) \stackrel{H}{=}10-2 \cdot 28-2 \cdot 37+2 \cdot 127-46+2^26-5^2 +145+4 \cdot 235 -3 \cdot 12^25
\\
&&
+2 \cdot 24^2+2 \cdot 3^24 -6 \cdot 1234-2^34 +3 \cdot 1^22^24,
\\
&&
g\left(3^3 4\right) \stackrel{H}{=} (12) -3 \cdot 2(10)-4 \cdot 39 +3 \cdot 129-48 +138 +3 \cdot 2^28
\\
&&
-3 \cdot 57 +15 \cdot 237 -9 \cdot 12^27
\\
&&
-3 \cdot 6^2 +3 \cdot 156 +12 \cdot 246+12 \cdot 3^26  -4 \cdot 2^36 -27 \cdot 1236 +9 \cdot 1^22^26
\\
&&
+3 \cdot 25^2  +9 \cdot 345-9 \cdot 1245 -9 \cdot 13^25 -15 \cdot 2^235  +9 \cdot 1^2235 +6 \cdot 12^35
\\
&&
-9 \cdot 2^24^2 -36 \cdot 23^24  +54 \cdot 12^234 +4 \cdot 2^44 -15 \cdot 1^22^34
\\
&&
-6 \cdot 3^4 +36 \cdot 123^3 +13 \cdot 2^33^2 -45 \cdot 1^22^23^2 -10 \cdot 12^43 +15 \cdot 1^32^33,
\\
&&
g\left(3^5\right) \stackrel{H}{=} (14)-5 \cdot 2(12)-5 \cdot 3(11) +5 \cdot 12(11) +10 \cdot 2^2(10)
\\
&&
-10 \cdot 59 +40 \cdot 239 -30 \cdot 12^29
\\
&&
-10 \cdot 68 +10 \cdot 158 +20 \cdot 248 +20 \cdot 3^28 -60 \cdot 1238 -20 \cdot 2^38 +30 \cdot 1^22^28
\\
&&
+30 \cdot 257 -90 \cdot 2^237 +60 \cdot 12^37
\\
&&
+30 \cdot 26^2 +60 \cdot 356 -90 \cdot 1256 -90 \cdot 2^246 -180 \cdot 23^26 +360 \cdot 12^236
\\
&&
+45 \cdot 2^46 -150 \cdot 1^22^36
\\
&&
+15 \cdot 45^2  -60 \cdot 3^35 -45 \cdot 135^2 -45 \cdot 2^25^2  +45 \cdot 1^225^2 -180 \cdot 2345 +180 \cdot 12^245
\\
&&
+360 \cdot 123^25 +210 \cdot 2^335  -450 \cdot 1^22^235  -150 \cdot 12^45 +150 \cdot 1^32^35
\\
&&
+60 \cdot 2^34^2 +360 \cdot 2^23^24 -600 \cdot 12^334 -51 \cdot 2^54 +225 \cdot 1^22^44
\\
&&
+ 120 \cdot 23^4 -600 \cdot 12^23^3 -225 \cdot 2^43^2 +900 \cdot 1^22^33^2 +315 \cdot 12^53 -525 \cdot 1^32^43
\\
&& +10 \cdot 2^7 -105 \cdot 1^22^6 +105 \cdot 1^42^5,
\\
&&
g(234) \stackrel{H}{=} 8-17 -2 \cdot 26 +1^26 -2 \cdot 35 +2 \cdot 125-4^2 +4 \cdot 134 +2 \cdot 2^24 -3 \cdot 1^224
\\
&&
+3 \cdot 23^2 -3 \cdot 1^23^2 -4 \cdot 12^23 +3 \cdot 1^323,
\\
&&
g\left(23^3\right) \stackrel{H}{=} (10)-19 -4 \cdot 28 +1^28-3 \cdot 37 +6 \cdot 127-3 \cdot 46
\\
&&
+6 \cdot 136 +9 \cdot 2^26 -9 \cdot 1^226
\\
&&
-3 \cdot 5^2 +6 \cdot 145 +21 \cdot 235 -9 \cdot 1^235 -24 \cdot 12^25 +9 \cdot 1^325 +6 \cdot 24^2 +6 \cdot 3^24 -36 \cdot 1234
\\
&&
-13 \cdot 2^34 +27 \cdot 1^22^24-6 \cdot 13^3 -27 \cdot 2^23^2 +36 \cdot 1^223^2 +55 \cdot 12^33 -45 \cdot 1^32^23
\\
&&
+4 \cdot 2^5 -25 \cdot 1^22^4 +15 \cdot 1^42^3,
\\
&&
g\left(2^2 3\right) \stackrel{H}{=} 6-2 \cdot 15 -3 \cdot 24 +3 \cdot 1^24 -3^2 +7 \cdot 123
\\
&&
-3 \cdot 1^33+2 \cdot 2^3 -7 \cdot 1^22^2 +3 \cdot 1^42.
\end{eqnarray*}
\end{small}
We shall not give the expressions for the 11 extra $(\pi)$ needed for $g_6$ when $l_3\neq 0$.

\section*{Appendix D: Expansions of $f_r,g_r$ in terms of ${\bf a}$}
\renewcommand{\theequation}{$\mbox{D.\arabic{equation}}$}
\setcounter{equation}{0}

Here, we give the coefficients $f(\pi)$, $g(\pi)$ needed in (\ref{fgpi}) for
$f_r$, $g_r$, $r\leq 4$ or 5 in terms of ${\bf a}$ of (\ref{ar}).
Again we first give these for the case  $l_3=0$.

For $r=2,3,4$,
$f(\pi)({\bf H})$ has fewer/the same/greater number of terms than
$f(\pi)({\bf a})$ in 19/4/5 cases,
and
$g(\pi)({\bf H})$ has fewer/the same/greater number of terms than
$g(\pi)({\bf a})$ in 11/2/2 cases.
On the other hand, as a function of $x$, $a_k$ is generally much briefer than $H_k$.
Here, we give the coefficients needed when $l_3=0$ up to $r=5$.
We again adapt the notation of Comtet (1974), but this time in terms of ${\bf a}$, not ${\bf H}$:
$k \cdot 1^{i_1}2^{i_2}\cdots  \stackrel{a}{=} ka_1^{i_1}a_2^{i_2}\cdots$.
For example, $15 \cdot 12^24^3  \stackrel{a}{=} 15a_1a_2^2a_4^3$.
MAPLE gave $f(\pi)$  in terms of ${\bf a}$ as follows:
\begin{small}
\begin{eqnarray*}
&&
r =2:\
f(4)=H_3 = a_1^3-3a_1a_2+a_3,
\
f(2)=H_1=a_1.
\\
&&
r =3:\ f(5) =H_4\mbox{ is given by (\ref{4a})},
\\
&&
f(14) =-3a_1^2a_2+3a_2^2+3a_1a_3 -a_4,
\\
&&
f(12) =-a_2.
\\
&&
r =4:\ f(6)=H_5\mbox{ is given by (\ref{5a})},
\\
&&
f\left(4^2\right) = -15a_1^5a_2 +96a_1^3a_2^2 -105a_1a_2^3
\\
&&
-204a_1^2a_2a_3 +105a_2^2a_3 +33a_1^4a_3  +69a_1a_3^2
\\
&&
-35a_1^3a_4 +105a_1a_2a_4 -35a_3a_4+21a_1^2a_5 -21a_2a_5-7a_1a_6 +a_7,
\\
&&
f(24) =-7a_1^3a_2+15a_1a_2^2+9a_1^2a_3 -10a_2a_3 -5a_1a_4 +a_5,
\\
&&
f\left(2^2\right) =-3a_1a_2 +a_3,
\\
&&
f(15) = -4a_1^3a_2 +12a_1a_2^2 +6a_1^2a_3  -10a_2a_3  -4a_1a_4 +a_5,
\\
&&
f\left(1^2 4\right) = 6a_1a_2^2+3a_1^2a_3-9a_2a_3-3a_1a_4+a_5,
\\
&&
f\left(1^2 2\right)  = a_3.
\end{eqnarray*}
 \end{small}
For $f_5$ when $l_3=0$, see Appendix D.

When $l_3=0$, the coefficients $g(\pi)$ in terms of ${\bf a}$
needed for $g_r(x)$ are as follows:
\begin{small}
\begin{eqnarray*}
&&
r =2:\ g(4)=H_3 \stackrel{a}{=} 1^3-3 \cdot 12+3,
\
g(2) =H_1=a_1  \stackrel{a}{=} 1.
\\
&&
r =3:\ g(5) =H_4\mbox{ of (\ref{4a})}.
\\
&&
r =4:\ g(6) =H_5\mbox{ of (\ref{5a})},
\\
&&
g\left(4^2\right)  \stackrel{a}{=}  -9 \cdot 1^52+72 \cdot 1^32^2-87 \cdot 12^3
\\
&&
+27 \cdot 1^43-180 \cdot 1^223+99 \cdot 2^23+63 \cdot 13^2+99 \cdot 124
\\
&&
-33 \cdot 34-33 \cdot 1^34+21 \cdot 1^25-21 \cdot 25-7 \cdot 16+7,
\\
&&
g(24)  \stackrel{a}{=} -3 \cdot 1^32 + 9 \cdot 12^2+6 \cdot 1^23-9 \cdot 23-4 \cdot 14+5,
\\
&&
g\left(2^2\right)   \stackrel{a}{=}  -12+3.
\\
&&
r =5:\ g(7) =H_6\mbox{ of (\ref{6a})},
\\
&&
g(45)  \stackrel{a}{=} -12 \cdot 1^62 +144 \cdot 1^42^2 -348 \cdot 1^22^3+96 \cdot 2^4
\\
&&
+42 \cdot 1^53 -480 \cdot 1^323 +774 \cdot 12^23
\\
&&
+258 \cdot 1^23^2 -270 \cdot 23^2
\\
&&
-64 \cdot 1^44 +396 \cdot 1^224 -204 \cdot 2^24 -268 \cdot 134 +34 \cdot 4^2+55 \cdot 1^35
\\
&&
-165 \cdot 125 +55 \cdot 35-28 \cdot 1^26 +28 \cdot 26+8 \cdot 17-8,
\\
&&
g(25) \stackrel{a}{=} -4 \cdot 1^42 +24 \cdot 1^22^2 -12 \cdot 2^3+10 \cdot 1^33
\\
&&
-46 \cdot 123 +10 \cdot 3^2-10 \cdot 1^24 +14 \cdot 24+5 \cdot 15-6.
\end{eqnarray*}
\end{small}
We skip $g_6$.
The extra terms needed for $f_2, \ldots, f_5,g_2, \ldots, g_5$ when $l_3\neq 3$ are given in Appendix D.
As functions of ${\bf a}$, the $c_r$ needed for $f_r$ in (\ref{ff}) are
\begin{eqnarray}
&&
c_2 =a_1,
\
c_3 = 2a_1^2+a_2,
\
c_4  = 3!a_1^3+7a_1a_2+a_3,
\nonumber
\\
&&
c_5 = 4!a_1^4+46a_1^2a_2 +11a_1a_3 +7a_2^2 +a_4,
\nonumber
\\
&&
c_6 = 5!a_1^5+ 324 a_1^3a_2 +147 a_1^2a_3 +127 a_1a_2^2 +16 a_1a_4 +25 a_2a_3+a_5.
\nonumber
\end{eqnarray}
So, $c_k$ has the same number of terms as a function of ${\bf H}$ or of ${\bf a}$.

We now give the extra terms needed when $l_3\neq 0$:
\begin{small}
\begin{eqnarray*}
&&
r =2:
\\
&&
f\left(3^2\right)   \stackrel{a}{=}  14 \cdot 12^2-8 \cdot 1^32+10 \cdot 1^23-10 \cdot 23-5 \cdot 14+5,
\\
&&
f(13)  \stackrel{a}{=} -2 \cdot 12+ 3.
\\
&&
r =3:
\\
&&
f(34)  \stackrel{a}{=} -11 \cdot 1^42+42 \cdot 1^22^2-15 \cdot 2^3 +19 \cdot 1^33 -59 \cdot 123
\\
&&
+10 \cdot 3^2 -15 \cdot 1^24 +15 \cdot 24 +2 \cdot 15 -6,
\\
&&
f\left(3^3\right)  \stackrel{a}{=}  138 \cdot 1^42^2 -374 \cdot 1^22^3 +104 \cdot 2^4
\\
&&
+26 \cdot 1^53 -500 \cdot 1^323 +810 \cdot 12^23 +280 \cdot 1^23^2 -280 \cdot 23^2
\\
&&
-280 \cdot 134-55 \cdot 1^44  +405 \cdot 1^224 -210 \cdot 2^24 +35 \cdot 4^2
\\
&&
+53 \cdot 1^35 -165 \cdot 125 +56 \cdot 35
\\
&&
-28 \cdot 1^26 +28 \cdot 26+8 \cdot 17 -8,
\\
&&
f(23)   \stackrel{a}{=} -5 \cdot 1^22+3 \cdot 2^2 +4 \cdot 13 -4,
\\
&&
f\left(13^2\right)   \stackrel{a}{=} -14 \cdot 2^3+24 \cdot 1^22^2+8 \cdot 1^33
\\
&&
-48 \cdot 123 +10 \cdot 3^2-10 \cdot 1^24 +15 \cdot 24+15-6,
\\
&&
f\left(1^2 3\right)  \stackrel{a}{=}  2 \cdot 2^2+2 \cdot 13-4.
\end{eqnarray*}
\end{small}
\begin{small}
\begin{eqnarray*}
&&
r =4:
\\
&&
f(35)   \stackrel{a}{=} -14 \cdot 1^52+96 \cdot 1^32^2-102 \cdot 12^3
\\
&&
+31 \cdot 1^43  -206 \cdot 1^223 +105 \cdot 2^23 +70 \cdot 13^2 -102 \cdot 3^3
\\
&&
+104 \cdot 124-34 \cdot 1^34 -35 \cdot 34+21 \cdot 1^25-21 \cdot 25-7 \cdot 16+7,
\\
&&
f\left(3^24\right)  \stackrel{a}{=}  222 \cdot 1^52^2-1094 \cdot 1^32^3+912 \cdot 12^4
\\
&&
-1053 \cdot 1^423 +35 \cdot 1^63+3615 \cdot 1^22^23
\\
&&
-2490 \cdot 123^2+810 \cdot 1^33^2-1259 \cdot 2^33+280 \cdot 3^3
\\
&&
-91 \cdot 1^54  +1185 \cdot 1^324 -1860 \cdot 12^24-1255 \cdot 1^234  +1260 \cdot 234 +315 \cdot 14^2
\\
&&
+121 \cdot 1^45 -749 \cdot 1^225+503 \cdot 135 +378 \cdot 2^25-126 \cdot 45
\\
&&
-82 \cdot 1^36 +250 \cdot 126-84 \cdot 36+36 \cdot 1^27-36 \cdot 27-9 \cdot 18+9,
\\
&&
f\left(3^4\right)  \stackrel{a}{=}   -3792 \cdot 1^52^3  + 14512 \cdot 1^32^4 - 9892 \cdot 12^5
\\
&&
- 1792 \cdot 1^623   + 27724 \cdot 1^42^23 + 3200 \cdot 1^53^2
\\
&&
-43360 \cdot 1^323^2 - 64976 \cdot 1^22^33  + 15400 \cdot 1^23^3 + 67910 \cdot 12^23^2
\\
&&
+ 17264 \cdot 2^43 - 15400 \cdot 23^3
\\
&&
- 80 \cdot 1^74- 31680 \cdot 1^32^24 + 4760 \cdot 1^524 - 10130 \cdot 1^434
\\
&&
+67880 \cdot 1^2234  + 5560 \cdot 1^34^2 - 17185 \cdot 124^2  + 33780 \cdot 12^34 - 23100 \cdot 13^24  - 34650 \cdot 2^234
\\
&&
+ 5775 \cdot 34^2 + 244 \cdot 1^65   - 5992 \cdot 1^425 +20028 \cdot 1^22^25 -27436 \cdot 1235  +8956 \cdot 1^335
\\
&&
- 6900 \cdot 1^245 - 6924 \cdot 2^35 + 6930 \cdot 245 + 4620 \cdot 3^25 + 1383 \cdot 15^2
\\
&&
- 350 \cdot 1^56 +4396 \cdot 1^326  - 6818 \cdot 12^26 - 4620 \cdot 1^236+ 2310 \cdot 146  + 4620 \cdot 236 - 462 \cdot 56
\\
&&
+ 298 \cdot 1^47   - 1980 \cdot 1^327   + 32 \cdot 1^227 + 1320 \cdot 137 + 990 \cdot 2^27 - 330 \cdot 47
\\
&&
+ 491 \cdot 128 - 161 \cdot 1^38   - 165 \cdot 38 + 55 \cdot 1^29 - 55 \cdot 29
\\
&&
- 11 \cdot 1(10) + (11),
\\
&&
f\left(23^2\right)   \stackrel{a}{=} 72 \cdot 1^32^2 -98 \cdot 12^3
\\
&&
+17 \cdot 1^43 -192 \cdot 1^223 +70 \cdot 13^2 +105 \cdot 2^23 -28 \cdot 1^34
\\
&&
+103 \cdot 124 -35 \cdot 34 +20 \cdot 1^25 -21 \cdot 25 -7 \cdot 16 +7,
\\
&&
f(134)  \stackrel{a}{=}  -84 \cdot 12^3-141 \cdot 1^223 +44 \cdot 1^32^2+11 \cdot 1^43 +59 \cdot 13^2 +104 \cdot 2^23 -19 \cdot 1^34
\\
&&
+89 \cdot 124-35 \cdot 34+19 \cdot 1^25-21 \cdot 25-6 \cdot 16 +7,
\\
&&
f\left(13^3\right)   \stackrel{a}{=} -552 \cdot 1^32^3  +748 \cdot 12^4
\\
&&
-406 \cdot 1^423 +2622 \cdot 1^22^23  +500 \cdot 1^33^2  -2180 \cdot 123^2
\\
&&
-1226 \cdot 2^33 +280 \cdot 3^3 -26 \cdot 1^54 +720 \cdot 1^324 -965 \cdot 1^234 -1620 \cdot 12^24 +280 \cdot 14^2
\\
&&
+1260 \cdot 234 +67 \cdot 1^45 -576 \cdot 1^225 +375 \cdot 2^25 +445 \cdot 135  -126 \cdot 45 -53 \cdot 1^36
\\
&&
+221 \cdot 126 -84 \cdot 36 +28 \cdot 1^27  -36 \cdot 27 -8 \cdot 18+ 9,
\\
&&
f(123)  \stackrel{a}{=}  10 \cdot 12^2 +5 \cdot 1^23 -10 \cdot 23 -4 \cdot 14+5,
\\
&&
f\left(1^23^2\right)   \stackrel{a}{=} -72 \cdot 1^223 -48 \cdot 12^3 +48 \cdot 13^2
\\
&&
+90 \cdot 2^23-8 \cdot 1^34+68 \cdot 124 -35 \cdot 34 +18 \cdot 1^25
\\
&&
-20 \cdot 25 -5 \cdot 16+ 7,
\\
&&
f\left(1^3 3\right) \stackrel{a}{=} -6 \cdot 23-2 \cdot 14+5.
\end{eqnarray*}
\end{small}
We now give $f_5$, but {\it only when $l_3=0$}, as the general case is too long to include here:
\begin{small}
\begin{eqnarray*}
&&
r =5:\ f(7)=H_6\mbox{ is given by (\ref{6a})},
\\
&&
f(45)   \stackrel{a}{=}
-19 \cdot 1^62 +189 \cdot 1^42^2 -411 \cdot 1^22^3 +105 \cdot 2^4
\\
&&
+51 \cdot 1^53 -542 \cdot 1^323 +276 \cdot 1^23^2 +837 \cdot 12^23-280 \cdot 23^2
\\
&&
-69 \cdot 1^44 +417 \cdot 1^224-210 \cdot 2^24 -279 \cdot 134 +35 \cdot 4^2
\\
&&
-168 \cdot 125 +56 \cdot 1^35+56 \cdot 35-28 \cdot 1^26+28 \cdot 26+8 \cdot 17 -8,
\\
&&
f(25)  \stackrel{a}{=} -9 \cdot 1^42 +42 \cdot 1^22^2 -15 \cdot 2^3+16 \cdot 1^33
\\
&&
-60 \cdot 123+10 \cdot 3^2-14 \cdot 1^24 +15 \cdot 24+2 \cdot 15 -6,
\\
&&
f(16)  \stackrel{a}{=}-5 \cdot 1^42+30 \cdot 1^22^2-15 \cdot 2^3+10 \cdot 1^33
\\
&&
-50 \cdot 123+10 \cdot 3^2-10 \cdot 1^24+15 \cdot 24+15 -6,
\\
&&
f\left(14^2\right)  \stackrel{a}{=} 75 \cdot 1^42^2 -288 \cdot 1^22^3 +105 \cdot 2^4
\\
&&
+15 \cdot 1^53 -324 \cdot 1^323 -33 \cdot 1^44+723 \cdot 12^23 +204 \cdot 1^23^2
\\
&&
-279 \cdot 23^2 +309 \cdot 1^224  -210 \cdot 2^24 -243 \cdot 134 +35 \cdot 4^2
\\
&&
+35 \cdot 1^35 -147 \cdot 125 +56 \cdot 35
\\
&&
-21 \cdot 1^26+28 \cdot 26+7 \cdot 17-8,
\\
&&
f(124) \stackrel{a}{=}21 \cdot 1^22^2  -15 \cdot 2^3+7 \cdot 1^33-48 \cdot 123
\\
&&
+10 \cdot 3^2-9 \cdot 1^24 +15 \cdot 24+15-6,
\\
&&
f\left(12^2\right)  \stackrel{a}{=} 3 \cdot 2^2+3 \cdot 13-4,
\\
&&
f\left(1^25\right)  \stackrel{a}{=}
12 \cdot 1^22^2-12 \cdot 2^3+4 \cdot 1^33 -36 \cdot 123 +10 \cdot 3^2-6 \cdot 1^24  +14 \cdot 24-6,
\\
&&
f\left(1^34\right)  \stackrel{a}{=} -6 \cdot 2^3
-18 \cdot 123 +9 \cdot 3^2-3 \cdot 1^24 +12 \cdot 24-15-6,
\\
&&
f\left(1^32\right)  \stackrel{a}{=} -4.
\end{eqnarray*}
\end{small}
We skip $f_6$.
The extra $g(\pi)$ needed for $g_r$,  $r=2,3,4$ in terms of ${\bf a}$ as follows:
\begin{small}
\begin{eqnarray*}
&&
r =2:
\\
&&
g\left(3^2\right)   \stackrel{a}{=}  -4 \cdot 1^32+10 \cdot 12^2+8 \cdot 1^23 -8 \cdot 23-5 \cdot 14+ 5.
\\
&&
r =3:
\\
&&
g(34)  \stackrel{a}{=} -6 \cdot 1^42+30 \cdot 1^22^2 -12 \cdot 2^3+15 \cdot 1^33
\\
&&
-51 \cdot 123+9 \cdot 3^2+14 \cdot 24-14 \cdot 1^24+6 \cdot 15-6,
\\
&&
g\left(3^3\right)  \stackrel{a}{=} 48 \cdot 1^42^2-212 \cdot 1^22^3+68 \cdot 2^4
\\
&&
-284 \cdot 1^323+8 \cdot 1^53+594 \cdot 12^23+226 \cdot 1^23^2-226 \cdot 23^2
\\
&&
-28 \cdot 1^44 +306 \cdot 1^224-168 \cdot 2^24-265 \cdot 134 +35 \cdot 4^2
\\
&&
-144 \cdot 125+38 \cdot 1^35+53 \cdot 35
\\
&&
-25 \cdot 1^26+25 \cdot 26+8 \cdot 17-8,
\\
&&
g(23) \stackrel{a}{=}-2 \cdot 1^22 +2 \cdot 2^2+3 \cdot 13 -4.
\end{eqnarray*}
\end{small}
\begin{small}
\begin{eqnarray*}
&&
r =4:
\\
&&
g(35)  \stackrel{a}{=} -8 \cdot 1^52+68 \cdot 1^32^2-84 \cdot 12^3
\\
&&
+24 \cdot 1^43-176 \cdot 1^223+66 \cdot 13^2+92 \cdot 2^23-30 \cdot 1^34
\\
&&
+98 \cdot 124-34 \cdot 34+20 \cdot 1^25-20 \cdot 25-7 \cdot 16+7,
\\
&&
g\left(3^24\right) \stackrel{a}{=}  84 \cdot 1^52^2-618 \cdot 1^32^3 +642 \cdot 12^4
\\
&&
-600 \cdot 1^423+2652 \cdot 1^22^23+12 \cdot 1^63
\\
&&
-2136 \cdot 123^2-1002 \cdot 2^33+624 \cdot 1^33^2 +252 \cdot 3^3
\\
&&
-48 \cdot 1^54+863 \cdot 1^324-1565 \cdot 12^24
\\
&&
-1126 \cdot 1^234+1141 \cdot 234+310 \cdot 14^2 +79 \cdot 1^45-629 \cdot 1^225+334 \cdot 2^25+483 \cdot 135-125 \cdot 45
\\
&&
-69 \cdot 1^36 +231 \cdot 126-81 \cdot 36 +34 \cdot 1^27-34 \cdot 27-9 \cdot 18+9,
\\
&&
g\left(3^4\right)  \stackrel{a}{=}  - 1008 \cdot 1^52^3  + 6752 \cdot 1^32^4 - 6044 \cdot 12^5
\\
&&
- 416 \cdot 1^623  + 12224 \cdot 1^42^23- 41344 \cdot 1^22^33 + 12124 \cdot 2^43 + 1264 \cdot 1^53^2
\\
&&
- 27728 \cdot 1^323^2+ 52552 \cdot 12^23^2 + 12896 \cdot 1^23^3  - 12896 \cdot 23^3
\\
&&
-16 \cdot 1^74  + 1840 \cdot 1^524 - 19320 \cdot 1^32^24 - 15395 \cdot 124^2 -5864 \cdot 1^434+ 53728 \cdot 1^2234
\\
&&
- 28524 \cdot 2^234 -21740 \cdot 13^24 + 25276 \cdot 12^34 +4200 \cdot 1^34^2 +5635 \cdot 34^2
\\
&&
+ 80 \cdot 1^65 - 3320 \cdot 1^425 + 14988 \cdot 1^22^25  - 5480 \cdot 2^35
\\
&&
- 24152 \cdot 1235 + 6760 \cdot 1^335 +4348 \cdot 3^25  - 6198 \cdot 1^245 + 6348 \cdot 245 + 1353 \cdot 15^2
\\
&&
- 168 \cdot 1^56  + 3136 \cdot 1^326 - 5704 \cdot 12^26- 4136 \cdot 1^236
\\
&&
+ 4136 \cdot 236 + 2280 \cdot 146 - 456 \cdot 56
\\
&&
+ 192 \cdot 1^47 - 1640 \cdot 1^227  + 852 \cdot 2^27 + 1288 \cdot 137- 330 \cdot 47
\\
&&
- 129 \cdot 1^38 + 451 \cdot 128 - 161 \cdot 38
\\
&&
+ 51 \cdot 1^29 - 51 \cdot 29  - 11 \cdot 1(10) +(11),
\\
&&
g\left(23^2\right)  \stackrel{a}{=}  20 \cdot 1^32^2 -50 \cdot 12^3+4 \cdot 1^43
\\
&&
-96 \cdot 1^223 +74 \cdot 2^23 +54 \cdot 13^2
\\
&&
-12 \cdot 1^34+73 \cdot 124 -33 \cdot 34+13 \cdot 1^25 -18 \cdot 25-6 \cdot 16 +7.
\end{eqnarray*}
\end{small}

\section*{Appendix E: Expansions of $f_r$, $g_r$ for the gamma when $l_3= 0$}
\renewcommand{\theequation}{$\mbox{E.\arabic{equation}}$}
\setcounter{equation}{0}

Here, we give the coefficients $f(\pi)$, $g(\pi)$ needed in (\ref{fgpi}) for
$f_r$, $g_r$, $r\leq 4$ or 5 when $l_3=0$ for $X=G$ a gamma variable with mean $m$
in terms of $\alpha=m-1$, $\overline{y}=-1/y$, as in (\ref{am}).

One can show
\begin{small}
\begin{eqnarray}
&&
f\left(1^i,k+2\right) = (-1)^i(k+1) \sigma^{k+i+1} \sum_{j=0}^k {k\choose j}
[\alpha]_{j+1} \overline{y}^{j+i+1}(j+2)_{i-1},
\
i\geq 1,
\label{fk1}
\\
&&
f\left( 1^i,2,k+1 \right) = (-1)^{i-1} \sigma^{k+i+2} \sum_{j=0}^k {k\choose j}
[\alpha]_{j} \overline{y}^{j+i+2} \left\{ (2k+1)(\alpha-j)+j^2\right\} (j+2)_{i}.
\label{fk2}
\end{eqnarray}
\end{small}
The other $f(\pi)$ are more simply given by the formulas for them in
Appendix C, but we give them here for the record:
\begin{small}
\begin{eqnarray*}
&&
r =2: \mbox{ for }k=4, 2, f(k)=H_{k-1}\mbox{ of (\ref{gee})}.
\\
&&
r =3:\ f(\pi)\mbox{ of (\ref{fk1}) for }\pi=5,12,14.
\\
&&
r =4:\
f(\pi)\mbox{ of (\ref{fk1}), (\ref{fk2}) for }\pi=6,15,24,2^2,1^24,1^22,25,
\\
&&
f\left(4^2\right)/ 3\sigma^7  = -5\alpha \overline{y}^2 - \alpha (25 \alpha-22) \overline{y}^3-
10  [\alpha]_2 (5 \alpha - 7) \overline{y}^4
\\
&&
-2 [\alpha]_2  \left(25 \alpha^2  - 89 \alpha + 84\right) \overline{y}^5
-[\alpha]_3 \left( 25 \alpha^2  - 109 \alpha + 140 \right) \overline{y}^6
\\
&&
-[\alpha]_3 \left(5\alpha^3 -39\alpha^2 +114 \alpha -120\right) \overline{y}^7.
\\
&&
r =5:\ f(\pi)\mbox{ of (\ref{fk1}), (\ref{fk2}) for }\pi=7,16,25,124,12^2,1^25,1^34,1^32,
\\
&&
f(45)/\sigma^8 =  - 19 \alpha \overline{y}^2  -6 \alpha (19 \alpha - 17) \overline{y}^3
-3 [\alpha]_2 (95 \alpha - 138) \overline{y}^4
\\
&&
-4 [\alpha]_2 \left(95 \alpha^2  - 349 \alpha + 336\right) \overline{y}^5
- 3 [\alpha]_3 \left(95 \alpha^2  - 433 \alpha + 560\right) \overline{y}^6
\\
&&
- 6 [\alpha]_3 \left(19 \alpha^3 -154 \alpha^2  +455 \alpha - 480\right)
\overline{y}^7
- [\alpha]_4 \left(19 \alpha^3 -177 \alpha^2  +638 \alpha - 840\right)
\overline{y}^8,
\\
&&
f\left(14^2\right)/3\sigma^8 =  10 \alpha \overline{y}^3  +3 \alpha (25 \alpha -22) \overline{y}^4
+40 [\alpha]_2 (5 \alpha -7) \overline{y}^5
\\
&&
+10 [\alpha]_2 \left(25 \alpha^2  - 89 \alpha + 84\right) \overline{y}^6
+6 [\alpha]_3 \left(25 \alpha^2  - 109 \alpha + 140\right) \overline{y}^7
\\
&&
+7 [\alpha]_3 \left(5 \alpha^3 -39 \alpha^2  +114 \alpha - 120\right) \overline{y}^8.
\end{eqnarray*}
\end{small}
\begin{small}
\begin{eqnarray*}
&&
r =6:
\
f(\pi)\mbox{ of (\ref{fk1}), (\ref{fk2})  for }\pi=8,17,
26,1^26,1^35,1^44,1^42,125,1^224,1^22^2,
\\
&&
f\left(5^2\right)/2\sigma^9 = -12 \alpha \overline{y}^2 -4 \alpha (21 \alpha-19) \overline{y}^3 -12 [\alpha]_2 (21 \alpha-31) \overline{y}^4
\\
&&
-12 [\alpha]_2 \left(35 \alpha^2 -130 \alpha +126\right) \overline{y}^5
\\
&&
-20 [\alpha]_3 \left(21 \alpha^2 -97 \alpha +126\right) \overline{y}^6
-12 [\alpha]_3 \left(21 \alpha^3 -172 \alpha^2 +511 \alpha -540\right) \overline{y}^7
\\
&&
-12 [\alpha]_4 \left(7 \alpha^3 -66 \alpha^2 +239 \alpha -315\right) \overline{y}^8
-4 [\alpha]_4 \left(3\alpha^4-43 \alpha^3 +258 \alpha^2 -743 \alpha +840\right) \overline{y}^9,
\\
&&
f(46)/\sigma^9 = -23 \alpha \overline{y}^2 - \alpha (161 \alpha-146) \overline{y}^3 -3 [\alpha]_2 (161 \alpha-242) \overline{y}^4
\\
&&
-5 [\alpha]_2 \left(161 \alpha^2 -610 \alpha +600\right) \overline{y}^5
\\
&&
-5 [\alpha]_3 \left(161 \alpha^2 -767 \alpha +1008\right) \overline{y}^6
\\
&&
-3 [\alpha]_3 \left(161 \alpha^3 -1357 \alpha^2 +4082 \alpha -4320\right) \overline{y}^7
\\
&&
- [\alpha]_4 \left(161 \alpha^3 -1575 \alpha^2 +5734 \alpha -7560\right)
\overline{y}^8
\\
&&
- [\alpha]_5 \left(2 3\alpha^3-249 \alpha^2 +1066 \alpha -1680\right)
\overline{y}^9,
\\
&&
f\left(4^3\right)/9\sigma^{11} = 14 \alpha \overline{y}^3 + 3 \alpha (57 \alpha-50) \overline{y}^4
\\
&&
+ \alpha \left(805 \alpha^2 -1860 \alpha +1064\right) \overline{y}^5
\\
&&
+ [\alpha]_2 \left(2023 \alpha^2 -6620 \alpha +5880\right) \overline{y}^6
\\
&&
+3 [\alpha]_2 \left(1015 \alpha^3 -5840 \alpha^2 +12056 \alpha -8720\right) \overline{y}^7
\\
&&
+ [\alpha]_2 \left(2849 \alpha^4  - 24931 \alpha^3  + 87918 \alpha^2  - 145044 \alpha + 92400\right) \overline{y}^8
\\
&&
+ [\alpha]_3 \left(1631 \alpha^4  - 16737 \alpha^3  + 71986 \alpha^2  - 149400 \alpha + 123200\right)
\overline{y}^9
\\
&&
+ [\alpha]_3 \left(525 \alpha^5  - 7500 \alpha^4  + 47457 \alpha^3  - 162474 \alpha^2  + 294472 \alpha - 221760\right)
\overline{y}^{10}
\\
&&
+ [\alpha]_3 \left(73 \alpha^6  - 1374 \alpha^5  + 11877 \alpha^4  - 59148 \alpha^3  + 175372 \alpha^2  - 288080 \alpha + 201600\right)
\overline{y}^{11},
\\
&&
f\left(24^2\right)/\sigma^9 = 62 \alpha \overline{y}^3 +81 \alpha (7 \alpha-6) \overline{y}^4
\\
&&
+3 \alpha  \left(635 \alpha^2  - 1450 \alpha +824\right) \overline{y}^5
\\
&&
+22 [\alpha]_2 \left(145 \alpha^2  - 473 \alpha + 420\right) \overline{y}^6
\\
&&
+36 [\alpha]_2 \left(80 \alpha^3  - 463 \alpha^2  + 961 \alpha - 700\right) \overline{y}^7
\\
&&
+3 [\alpha]_3  \left(449 \alpha^3 - 3079 \alpha^2  + 8006 \alpha - 7560\right) \overline{y}^8
\\
&&
+ [\alpha]_3  \left(257 \alpha^4  - 2691 \alpha^3  + 11746 \alpha^2  - 24552 \alpha + 20160\right) \overline{y}^9,
\\
&&
f\left(2^2 4\right)/3\sigma^7 = 10 \alpha \overline{y}^3 + \alpha (61 \alpha-50) \overline{y}^4
\\
&&
+ \alpha  \left(123 \alpha^2  - 272 \alpha + 152\right) \overline{y}^5
\\
&&
+ [\alpha]_2 \left(103 \alpha^2  - 326 \alpha + 280\right)
\overline{y}^6
\\
&&
+ [\alpha]_3 \left(31 \alpha^2  - 114 \alpha + 120\right) \overline{y}^7,
\\
&&
f\left(2^3\right)/\sigma^5 = 14 \alpha \overline{y}^3 + \alpha (43 \alpha-30) \overline{y}^4
\\
&&
+ \alpha  \left(29 \alpha^2  - 50 \alpha + 24 \right) \overline{y}^5,
\\
&&
f(145)/2\sigma^9 = 19 \alpha \overline{y}^3 + 9 \alpha (19 \alpha-17) \overline{y}^4
 + 6 [\alpha]_2  (95 \alpha -138) \overline{y}^5
\\
&&
+10 [\alpha]_2 \left( 95 \alpha^2  - 349 \alpha + 336 \right) \overline{y}^6
\\
&&
+9 [\alpha]_3 \left(95 \alpha^2  - 433 \alpha + 560\right) \overline{y}^7
\\
&&
+21 [\alpha]_3  \left( 19 \alpha^3  - 154 \alpha^2  + 455 \alpha - 480 \right) \overline{y}^8
\\
&&
+4 [\alpha]_4  \left(19 \alpha^3  - 177 \alpha^2  + 638 \alpha - 840\right) \overline{y}^9,
\\
&&
f\left(1^2 4^2\right)/6\sigma^9 =  -15 \alpha \overline{y}^4 -6 \alpha (25 \alpha-22) \overline{y}^5 -100 [\alpha]_2 (5\alpha-7) \overline{y}^6
\\
&&
-30 [\alpha]_2  \left(25 \alpha^2  - 89 \alpha + 84 \right) \overline{y}^7
\\
&&
-21 [\alpha]_3  \left(25 \alpha^2  - 109 \alpha + 140\right)\overline{y}^8
\\
&&
-28 [\alpha]_3  \left(5 \alpha^3  - 39 \alpha^2  + 114 \alpha - 120\right) \overline{y}^9.
\end{eqnarray*}
\end{small}
When $l_3=0$, $ g(\pi)$  needed for $g_r$ in  (\ref{fgpi}) are as follows:
\begin{small}
\begin{eqnarray}
&&
g(2,k+1) = -\sigma^{k+2} \sum_{j=0}^k {k\choose j} [\alpha]_{j+1} \left\{ k\alpha-(k-1)j \right\} \overline{y}^{j+2}.
\label{ggeek}
\\
&&
r =2: \mbox{ for }k=4,2,\ g(k)=H_{k-1}\mbox{ of (\ref{gee}).}
\nonumber
\\
&&
r =3:\ g(5)=H_{4}\mbox{ of (\ref{gee}).}
\nonumber
\\
&&
r =4:\ g(6)=H_{5}\mbox{ of (\ref{gee}) and }
g(24), g \left(2^2\right) \mbox{ of (\ref{ggeek}),}
\nonumber
\\
&&
g\left(4^2\right)/ 3\sigma^7  = -3\alpha \overline{y}^2 -3 \alpha (5 \alpha-6) \overline{y}^3
-6  [\alpha]_2 (5 \alpha - 11) \overline{y}^4
\nonumber
\\
&&
-6  [\alpha]_2  \left(5 \alpha^2  - 25 \alpha + 28\right) \overline{y}^5
-[\alpha]_3 \left(15 \alpha^2  - 99 \alpha + 140\right) \overline{y}^6
\nonumber
\\
&&
-[\alpha]_3 (\alpha - 6) \left( 3 \alpha^2  - 15 \alpha + 20 \right) \overline{y}^7.
\nonumber
\\
&&
r = 5:\ g(7)=H_{6}\mbox{ of (\ref{gee})},
\
g(25)\mbox{ of (\ref{ggeek})},
\nonumber
\\
&&
g(45)/12\sigma^8 =  -\alpha \overline{y}^2  - \alpha (6 \alpha - 7) \overline{y}^3 - [\alpha]_2 (15 \alpha - 32) \overline{y}^4
\nonumber
\\
&&
-2 [\alpha]_2 \left( 10 \alpha^2  - 49 \alpha + 55 \right) \overline{y}^5
-[\alpha]_3 \left(15 \alpha^2  - 97 \alpha + 140\right) \overline{y}^6
\nonumber
\\
&&
- [\alpha]_3 \left( 6 \alpha^3  - 65 \alpha^2  + 219 \alpha - 240 \right)
\overline{y}^7 -  [\alpha]_4 (\alpha - 7) \left(\alpha^2  - 6 \alpha + 10\right) \overline{y}^8.
\nonumber
\end{eqnarray}
\end{small}
\begin{small}
\begin{eqnarray*}
&&
r =6:\ g(8)=H_{7}\mbox{ of (\ref{gee})},
\
g(26)\mbox{ of (\ref{ggeek})},
\\
&&
g\left(5^2\right)/8\sigma^9 = -2\alpha \overline{y}^2  -2\alpha(7 \alpha-8) \overline{y}^3 -
3 [\alpha]_2  ( 14  \alpha -29) \overline{y}^4
- [\alpha]_2  \left( 70 \alpha^2- 335  \alpha + 372\right) \overline{y}^5
\\
&&
- 10[\alpha]_3 \left( 7 \alpha^2  - 44  \alpha + 63\right) \overline{y}^6
- 6 [\alpha]_3 \left( 7 \alpha^3 -74 \alpha^2 +247  \alpha -270\right) \overline{y}^7
\\
&&
-  [\alpha]_4 \left(14 \alpha^3  - 177 \alpha^2  + 703 \alpha - 945\right) \overline{y}^8 -
[\alpha]_4 (\alpha - 8) (2 \alpha - 7) \left(\alpha^2 - 7 \alpha + 15\right) \overline{y}^9,
\\
&&
g(46)/15 \sigma^9 = -\alpha \overline{y}^2  - \alpha (7 \alpha - 8) \overline{y}^3
 - [\alpha]_2 (21 \alpha - 44) \overline{y}^4
\\
&&
- [\alpha]_2 \left(35 \alpha^2  - 170 \alpha + 192\right) \overline{y}^5
- [\alpha]_3 \left(35 \alpha^2 - 225 \alpha + 332\right) \overline{y}^6
\\
&&
- [\alpha]_3  \left(21 \alpha^3  - 227 \alpha^2 + 776 \alpha - 864\right) \overline{y}^7
- [\alpha]_4  \left(7 \alpha^3  - 91 \alpha^2  + 372 \alpha - 504\right) \overline{y}^8
\\
&&
- [\alpha]_5  (\alpha - 8) \left(\alpha^2 - 7 \alpha + 14\right) \overline{y}^9,
\\
&&
g\left(4^3\right)/9\sigma^{11} = 6   \alpha \overline{y}^3 + 12 \alpha   ( 5 \alpha-6) \overline{y}^4
+ 3   \alpha \left( 119 \alpha^2  - 380 \alpha +264\right) \overline{y}^5
\\
&&
+ 3 [\alpha]_2  \left(301 \alpha^2  - 1480 \alpha + 1680\right) \overline{y}^6
+3 [\alpha]_2 \left(455 \alpha^3  - 3800 \alpha^2  + 9864 \alpha - 8160\right) \overline{y}^7
\\
&&
+3 [\alpha]_2 \left(427 \alpha^4  - 5333 \alpha^3  + 23454 \alpha^2  - 44092 \alpha + 30240\right) \overline{y}^8
\\
&&
+ [\alpha]_3 \left(735 \alpha^4  - 11265 \alpha^3  + 60786 \alpha^2  - 141816 \alpha + 123200\right) \overline{y}^9
\\
&&
+ [\alpha]_3 \left(237 \alpha^5   - 4944 \alpha^4  + 39033 \alpha^3  - 150558 \alpha^2  + 288712 \alpha - 221760\right) \overline{y}^{10}
\\
&&
+ [\alpha]_3  (\alpha - 10) \left(33 \alpha^5  - 564 \alpha^4  + 3957 \alpha^3  - 14298 \alpha^2  + 26552 \alpha - 20160\right) \overline{y}^{11},
\\
&&
g\left(24^2\right)/3\sigma^9 = 6 \alpha \overline{y}^3  + 3 \alpha  (19 \alpha - 24)
\overline{y}^4 + 3 \alpha  \left(65 \alpha^2  - 222 \alpha+160\right) \overline{y}^5
\\
&&
+6 [\alpha]_2 \left(55 \alpha^2  - 297 \alpha + 360\right) \overline{y}^6
+12 [\alpha]_2  \left(25 \alpha^3  - 230 \alpha^2  + 642 \alpha - 560\right) \overline{y}^7
\\
&&
 + [\alpha]_3 \left(141 \alpha^3  - 1665 \alpha^2  + 5948 \alpha - 6720\right) \overline{y}^8
\\
&&
+ [\alpha]_3 (\alpha - 8) (3 \alpha - 10) \left(9 \alpha^2  - 53 \alpha + 84\right) \overline{y}^9,
\\
&&
g\left(2^2 4\right)/3\sigma^7 = 2\alpha \overline{y}^3 +\alpha  (13 \alpha - 18) \overline{y}^4  +
\alpha  \left(27 \alpha^2  - 104 \alpha+ 80\right) \overline{y}^5
\\
&&
+ [\alpha]_3  (23 \alpha - 100) \overline{y}^6
+ [\alpha]_3 (\alpha - 6) (7 \alpha - 20) \overline{y}^7,
\\
&&
g\left(2^3\right)/\sigma^5 = 2 \alpha \overline{y}^3 +\alpha  (7 \alpha - 12) \overline{y}^4  +
\alpha  (\alpha - 4) (5 \alpha - 6) \overline{y}^5.
\end{eqnarray*}
\end{small}

\section*{Appendix F:  Comparison with Cornish and Fisher}
\renewcommand{\theequation}{$\mbox{F.\arabic{equation}}$}
\setcounter{equation}{0}

Here, we specialize to the case $X=N$,
so that $H_r=He_r$,  the $r$th Hermite polynomial of (\ref{He}).
We give $f(\pi)$, $g(\pi)$ of (\ref{fgpi}) needed for $f_r$, $g_r$ of
(\ref{fff}) and (\ref{ggg}).
We also give special formulas for some of them that do not hold in the non-normal case.
Recall that for $i\geq 2$, $\mathbb{E}\ N^{2j}=1 \cdot 3\cdots (2j-1)$.
Using the notation (\ref{[]j}), we have
\begin{eqnarray}
f\left(1^k,k\right) =0,
\
f\left(1^j,k+1\right) =(-1)^j [k]_j\ H_{k-j},
\
f\left(2^j\right) = (-1)^{j-1}H_1\ \mathbb{E} N^{2j}.
\nonumber
\end{eqnarray}
As above, special relations are put in the line starting '$r=$'.
$f(\pi)$ needed for $f_r$ are as follows:
\begin{small}
\begin{eqnarray*}
&&
r =1:\  f(3)=H_2,
\
f(1)=1.
\\
&&
r =2:\  f(4)=H_3,
\
f(2)=H_1,
\
f(13) =- 2H_1,
\\
&&
f\left(3^2\right) = -2\left(4x^3-7x\right).
\\
&&
r =3:\ f(5) = H_4,
\
f(14)  = -3H_2,
\
f(12) = -1,
\
f\left(1^23\right) =2,
\\
&&
f(34) = -11x^4+42x^3-15,
\
f\left(3^3\right) = 2\left(69x^4-187x^3+52\right),
\\
&&
f(23)  =-5x^2+3,
\
f\left(13^2\right) = 2\left(12x^2-7\right).
\\
&&
r =4: \ f(6)  = H_5,
\
f\left(2^2\right) =  -3H_1,
\
f(15) = -4H_3,
\
f\left(1^2 4\right)=6H_1,
\\
&&
f\left(1^2 2\right) = f\left(1^3 3\right) =0,
\\
&&
f\left(4^2\right)  =  -3\left(5 x^5-32 x^3+35 x\right),
\
f(35)  = -2\left(7 x^5-48 x^3+51 x\right),
\\
&&
f\left(3^2 4\right) = 2\left(111x^5-547x^3+456x\right),
\
f\left(3^4\right)  =  -4\left(948 x^5-3628 x^3+2473 x\right),
\\
&&
f(24)  = -\left(7x^3-15x\right),
\
f\left(23^2\right)  = 2\left(36x^3-49x\right),
\
f(134) =4\left(11 x^3- 21 x\right),
\\
&&
f\left(1 3^3\right) = -4\left(138 x^3- 187 x\right),
\
f(123)=10H_1,
\
f\left(1^2 3^2\right) =-48H_1.
\end{eqnarray*}
\begin{eqnarray*}
&&
r =5: \ f(7)  = H_6,
\
f(16)  =-5H_4,
\
f\left(1^2 5\right) = 12H_2,
\
f\left(1^34\right) = -6,
\\
&&
f\left(1^3 2\right) =f\left(1^4 3\right)=0,
\\
&&
f(45)  = -19x^6 +189x^4  -411x^2+105,
\
f(36)  = -17x^6 +185x^4  -405x^2+105,
\\
&&
f\left(3 4^2\right)  = 347 x^6 -2643 x^4 +4521 x^2 -945,
\
f\left(3^2 5\right)  = 2\left(162 x^6 -1309 x^4 +2232 x^2 -471\right),
\\
&&
f\left(3^3 4\right) = -2\left( 3354 x^6 -20831 x^4 +29148 x^2 -5174\right),
\\
&&
f\left(3^5\right)  = 4 \left(36240 x^6- 184146 x^4 + 217921 x^2 -33523\right),
\\
&&
f(25)  =  -3\left(3x^4-14 x^2 +5\right),
\
f(234)  = 121  x^4 -378 x^2 +105,
\\
&&
f\left(2 3^3\right)  = -2\left( 897 x^4 -2057 x^2 + 468 \right),
\
f\left(2^2 3\right) = 5\left(7 x^2-3\right),
\\
&&
f\left(1 4^2\right) =  3\left(25 x^4 -96 x^2 +35\right),
\
f(135)  = 2\left(35 x^4 -144 x^2 + 51\right),
\\
&&
f\left(1 3^2 4\right)  = -6\left(185 x^4 -547 x^2 +152\right),
\
f\left(1 3^4\right)  = 4\left(4740 x^4 -10884 x^2 + 2473\right),
\\
&&
f(124) = 3\left(7 x^2-5\right),
\
f\left(123^2\right) = -2\left(108 x^2-49\right),
\
f\left(1 2^2\right) = 3,
\\
&&
f\left(1^2 34\right) = -12 \left(11x^2-7\right),
\
f\left(1^2 3^3\right) = 4\left(414 x^2-187\right),
\
f\left(1^2 23\right) = -10,
\
f\left(1^3 3^2\right) = 48.
\end{eqnarray*}
 \end{small}
$f_6$ has forty six terms, so is only given here for the case $l_3=0$:
\begin{small}
\begin{eqnarray*}
&&
r =6: \ f(8)  = H_7,
\
f(17) = -6H_5,
\
f\left(1^26\right) = 20 H_3,
\
f\left(1^35\right) = -24 H_1,
\
f\left(2^3\right) = 15H_1,
\\
&&
f\left(1^33\right) = f\left(1^44\right) = f \left(1^53\right) = f\left(1^42\right)=f\left(1^43^2\right)=f\left(1^323\right)=0,
\\
&&
f\left(5^2\right) = -24 \left(x^7 -14 x^5 +51 x^3 -39 x\right),
\
f(46)  = -23 x^7 +333 x^5  -1215 x^3+945 x,
\\
&&
f\left(4^3\right) = 3 \left(177 x^7 -1899 x^5 +5451 x^3 -3465 x\right),
\
f(26) =  -11 x^5 + 90 x^3  - 105 x,
\\
&&
f\left(24^2\right) =  3\left( 65 x^5 -352 x^3  +315 x\right),
\
f\left(2^24\right) = 21\left(3 x^3 -5 x\right),
\\
&&
f(145) = 6\left(19x^5-126x^3+137x\right),
\
f(125) = 12\left(3x^3-7x\right),
\\
&&
f\left(1^24^2\right) = -12\left(25 x^3  -48 x\right),
\
f\left(1^224\right) =  -42 H_1.
\end{eqnarray*}
\end{small}
Note the special relations
\begin{eqnarray}
f\left(1^2 3^2\right)  =-48H_1,
\
f\left(1^224\right) =  -42 H_1,
\
f(123)=10H_1.
\nonumber
\end{eqnarray}
For $k\geq 2$, $i\geq 0$, $g(2^i,k) = (-1)^i\nu_{ki}H_{k-1}$,
where $\nu_{k0}=1$, $\nu_{ki}=(k-1)(k+1)\cdots (k+2i-3)$ for $i\geq 1$.
That is,
\begin{eqnarray*}
g\left( 2^i,r+2-2i \right) =(-1)^i <r>_i H_{r+1-2i},
\
0\leq i\leq r/2,
\end{eqnarray*}
where
\begin{eqnarray*}
&&
<r>_0=1,
\\
&&
<r>_i = (r-1)(r-3)\cdots (r-2i+1) =2^r \Gamma(s)/ \Gamma(s-i),
\
s=(r+1)/2,
\
i\geq 1.
\end{eqnarray*}
Again, cases of this relation are put in the line next to '$r=$'.
The coefficients  $g(\pi)$ of (\ref{fgpi}) needed for $g_r$ are as follows:
\begin{small}
\begin{eqnarray*}
&&
r =1:\  g(3)=H_2,
\
g(1)=1.
\\
&&
r =2:\ g(4)=H_3,
\
g(2)=H_1,
\\
&&
g\left(3^2\right) =-2\left(2x^3-5x\right).
\\
&&
r =3:\   g(5) = H_4,
\
g(23)  = -2H_2,
\\
&&
g(34) = -6\left(x^4-5x^2+2\right),
\
g\left(3^3\right) = 4\left(12x^4-53x^2+17\right).
\\
&&
r =4: \  g(6)  = H_5,
\
g(24) = -3H_3,
\
g\left(2^2\right) = -H_1,
\\
&&
g\left(4^2\right) =-3\left(3x^5-24x^3+29x\right),
\
g(35)  = -4\left(2x^5-17x^3+21x\right),
\\
&&
g\left(3^2 4\right) = 6\left(14x^5-103x^3+107x\right),
\
g\left(3^4\right)  = -4\left(252x^5-1688x^3+1511x\right),
\\
&&
g\left(23^2\right) = -5g\left(3^2\right).
\\
&&
r =5:\ g(7)=H_6,
\
g(25) =-4H_4,
\
g\left(2^2 3\right) =8H_2,
\\
&&
g(45) =-12\left(x^6-12x^4+29x^2-8\right),
\
g(36)  =-10\left(x^6-13x^4+33x^2-9\right),
\\
&&
g\left(34^2\right) = 12\left(12x^6-129x^4+271x^2-64\right),
\
g\left(3^25\right) =8\left(16x^6-181x^4+393x^2-90\right),
\\
&&
g\left(3^3 4\right) = -24\left(80x^6-803x^4+1513x^2-304\right),
\\
&&
g\left(3^5\right)  = 32\left(960x^6-8937x^4+15062x^2-2651\right),
\\
&&
g(234) = -6g(34),
\
g\left(2 3^3\right) = -8g\left(3^3\right).
\end{eqnarray*}
\end{small}
\begin{small}
\begin{eqnarray*}
&&
r =6:\ g(8) =H_7,
\
g(26) = -5H_5,
\
g\left(2^2 4\right)  = 15H_3,
\
g\left(2^3\right)  = 3H_1,
\\
&&
g\left(5^2\right) =-8\left(2x^7-33x^5+132x^3-108x\right),
\\
&&
g(46)  = -15\left(x^7-17x^5+69x^3-57x\right),
\\
&&
g\left(4^3\right) = 27\left(9x^7-131x^5+451x^3-321x\right),
\\
&&
g(37) = 6\left(2x^7-37x^5+160x^3-135x\right),
\\
&&
g(345) = 12\left(18x^7-273x^5+974x^3-695x\right),
\\
&&
g\left(3^2 6\right) = 10\left(18x^7-293x^5+1100x^3-795x\right),
\\
&&
g\left(3^2 4^2\right) = -6\left(594x^7-8193x^5+26006x^3-16367x\right),
\\
&&
g\left(3^3 5\right)  = -8\left(396x^7-5708x^5+18755x^3-11811x\right),
\\
&&
g\left(3^44\right) = 12\left(5148x^7-67004x^5 +195259 x^3 -109553 x\right),
\\
&&
g\left(3^6\right) = -8\left( 154440x^7-1887684x^5+5033714x^3-2542637x \right),
\\
&&
g\left(2 4^2\right) = -7g\left(4^2\right),
\
g(235) = -7g(35),
\
g\left(23^2 4\right)  = -9g \left(3^2 4\right),
\\
&&
g\left(2 3^4\right)  = -11g\left(3^4\right),
\
g\left(2^2 3^2\right)= 35 g \left( 3^2 \right).
\end{eqnarray*}
\end{small}
These results agree with those given on page 317 of Cornish and Fisher (1937)
and page 214 of Fisher and Cornish (1960), except for
(i) a typo on page 316: $ab$ was written for $ad$,
making $f(14)$ above into a second $f(12)$; and
(ii) they gave $f(15)=-4(x^3-x)$ instead of $-4H_3$.
The factor $-6$ in $g(34)$ was omitted by Hill and Davis (1968).
$x$ in $f(3^4)$  was omitted in Withers (1984).
Fisher and Cornish did not give $h_r$, $f_r$ for $r=5,6$.
Hill and Davis reference a Stanford Technical Report which may give $e_r(x,L)$
for $e=h$, $f$, $g$ up to $r=11$ when $X=N$.

Note how  $g(2\pi)$ is a multiple of $g(\pi)$.
For example, if $\pi=(3^{i_3}4^{i_4}\cdots)$, then $g(2\pi)=(1-|\pi|)g(\pi)$, where $|\pi|=\sum_{k=3} ki_k$.

\section*{Acknowledgments}

The authors would like to thank the Editor and the referee
for careful reading and for their comments which greatly improved the paper.

\newpage

\begin{small}
{\bf Table 1.1}~~Number of terms needed by $e_{r}(x)$ for different choices of $X$.\\
\begin{tabular}{|c|c|c|c|c|c|c|c|c|c|c|c|}
\hline
$J$, $K$:  &  &${0,1}$&${1,2}$&${JK}$&J&K& &${01}$&${12}$&${JK}$& saving
\\
\hline
& \multicolumn{4}{|c|}{Number of terms,  $N+M$} &&
& \multicolumn{4}{|c|}{Cumulative number of terms,  $N+M$} &
\\
\hline
$e_{0}$& 1+ 0& 1+ 0& 1+ 0& 1+ 0& 0 & 1& 1+ 0& 1+ 0& 1+ 0& 1+ 0& 0\%\\
$e_{1}$& 2+ 0& ``  & 0+ 0& 0+ 0& 1 & 1& 3+ 0& 2+ 0 & `` & ``  & 67\%\\
$h_{2}$& 5+ 0& 3+ 0& 1+ 0& 1+ 0& 1 & 2& 8+ 0& 5+ 0& 2+ 0& 2+ 0& 75\%\\
$f_{2}$& 3+ 0& 2+ 0& ``  & ``  & 1 & 2& 6+ 0& 4+ 0& ``  & ``  & 67\%\\
$g_{2}$& ``  & ``  & ``  & ``  & 1 & 2& ``  & ``  & ``  & ``  & 67\%\\
$h_{3}$& 9+ 2& 4+ 2& 1+ 2& 1+ 1& 2 & 2&17+ 2& 9+ 2& 3+ 2& 3+ 1& 79\%\\
$f_{3}$& 8+ 2& 3+ 2& ``  & ``  & 2 & 2&14+ 2& 7+ 2& ``  & ``  & 75\%\\
$g_{3}$& 4+ 2& 1+ 2& ``  & ``  & 2 & 2&10+ 2& 5+ 2& ``  & ``  & 67\%\\
$h_{4}$&17+ 6& 8+ 4& 2+ 2& 2+ 1& 2 & 3&34+ 8&17+ 6& 5+ 4& 5+ 2& 83\%\\
$f_{4}$&14+ 5& 7+ 4& ``  & ``  & 2 & 3&28+ 7&14+ 6& ``  & ``  & 80\%\\
$g_{4}$& 8+ 3& 4+ 2& ``  & ``  & 2 & 3&18+ 5& 9+ 4& ``  & ``  & 70\%\\
$h_{5}$&28+15&11+10& 2+ 5& 2+ 2& 3 & 3&62+23&28+16& 7+ 9& 7+ 4& 87\%\\
$f_{5}$&25+15&10+10& ``  & ``  & 3 & 3&53+22&24+16& ``  & ``  & 85\%\\
$g_{5}$&11+ 8& 3+ 5& 2+ 4& ``  & 3 & 3&29+13&12+ 9& 7+ 8& ``  & 74\%\\
$h_{6}$&46+42&19+32& 4+10& 4+ 3& 3 & 4&108+65&47+48&11+19&11+ 7& 90\%\\
$f_{6}$&40+37&18+22& 4+ 9& ``  & 3 & 4&93+59&42+38&11+18& ``   & 88\%\\
$g_{6}$&19+16& 8+ 9& 4+ 7& ``  & 3 & 4&48+29&20+18&11+15& ``   & 77\%\\
\hline
\end{tabular}
\end{small}

\begin{small}
{\bf Table 2.1}~~Two approximations for the 95th quantile of $2^{-1}\ln F_{24,60}$.
\\
\begin{tabular}{|c|c|c|c|}
\hline
Order of & Successive & Successive & Successive \\
magnitude & terms & totals & errors \\
\hline
0 &  .2809 1224 & .2809 1224 &  .0155 6380 \\
1 & -.0196 0643 & .2613 0581 & -.0040 4263 \\
2 &  .0044 6851 & .2657 7432 &  .0004 2588 \\
3 & -.0004 8004 & .2652 9428 & -.0000 5416 \\
4 &  .0000 5645 & .2653 5073 &  .0000 0229 \\
5 & -.0000 0154 & .2653 4919 & -.0000 0075 \\
6 & -.0000 0102 & .2653 4817 &  .0000 0027 \\
\hline
\end{tabular}
\end{small}

\begin{small}
{\bf Table 2.2}~~Two approximations for the 95 quantile of $2^{-1}\ln F_{5,5}$.
\\
\begin{tabular}{|c|c|c|c|}
\hline
Order of & Successive & Successive & Successive  \\
magnitude & terms & totals & errors \\
\hline
0 &  .0 & . &  .   \\
1 & -.0 & . & -.00 \\
2 &  .0 & . &  .00 \\
3 & -.0 & . & -.00 \\
4 &  .0 & . &  .00 \\
5 & -.0 & . & -.00 \\
6 & -.0 & . &  .00 \\
\hline
\end{tabular}
\end{small}

\end{document}